\pdfoutput=1
\documentclass[10pt]{article}
\usepackage{wrapfig,cancel, mathrsfs}
\usepackage{color}

\usepackage{graphicx, framed}
\usepackage[colorlinks=true, pdfstartview=FitV, linkcolor=blue, citecolor=blue,urlcolor=blue]{hyperref}
\definecolor{shadecolor}{rgb}{0.95, 0.95, 0.86}

\definecolor{gray}{rgb}{0.5,0.5,0.5}

\def \Id{ {\mathrm {Id}}}

\usepackage{amsbsy}
\usepackage{amssymb}
\newtheorem{problem}{Problem}[section]
\def\D{\mathbb D}
\newtheorem{question}{Question}

\def\QED{ {\bf Q.E.D.}}

\usepackage{verbatim}
\def\tr {\mathrm {Tr}}

\renewcommand{\theequation}{\arabic{section}.\arabic{equation}}

\makeatletter
\@addtoreset{equation}{section}
\makeatother

\def\b{\vec{b}}
\def\a{\vec{a}}
\def\bR{{\mathbb R}}
\def\F{{\mathcal F}}
\def\vtau{{\vec\tau}}
\def\I{{\vec I}}

\def\Ai{\mathrm{Ai}}
\def\t{{\vec t}}

\newtheorem{theorem}{Theorem}[section]
\newtheorem{example}{Example}[section]
\newtheorem{exercise}{Exercise}[section]

\newtheorem{lemma}{Lemma}[section]
\newtheorem{remark}{Remark}[section]

\newtheorem{proposition}{Proposition}[section]
\newtheorem{corollary}{Corollary}[section]
\newtheorem{definition}{Definition}[section]
\def\le{\left}
\def\ri{\right}
\def\ds{\displaystyle}

\def\eqref#1{(\ref{#1})}

\def\un{\underline}
\def\res{\mathop{\mathrm {res}}\limits}

\def\S {\mathbf S}
\def\br{\begin{remark}}
\def\er{\end{remark}}
\def\bt{\begin{theorem}}
\def\et{\end{theorem}}
\def \s{{\vec s}}
\def\bc{\begin{corollary}}
\def\ec{\end{corollary}}
\def\bx{\begin{example}\small}
\def\ex{\end{example}}
\def\bxr{\begin{exercise}\small}
\def\exr{\end{exercise}}
\def\bl{\begin{lemma}}
\def\el{\end{lemma}}

\def\bd{\begin{definition}}
\def\ed{\end{definition}}
\def\bp{\begin{proposition}}
\def\ep{\end{proposition}}
\def\be{\begin{equation}}
\def\ee{\end{equation}}

\def\bes{$$}
\def\ees{$$}
\def\&{\hspace{-15pt}&}
\def\bea{\begin{eqnarray}}
\def\eea{\end{eqnarray}}
\def\beas{\begin{eqnarray*}}
\def\eeas{\end{eqnarray*}}

\def \pa{\partial}
\def\C{{\mathbb C}}

\def\R{{\mathbb R}}

\def\z {{\zeta}}

\def\H{{\cal H}}

\def\d{\,\mathrm d}

\def\1{{\bf 1}}
\def\wt{\widetilde}

\date{}
\begin{document}

\newpage
\baselineskip 15pt plus 1pt minus 1pt

\vspace{0.2cm}
\begin{center}
\begin{Large}
\fontfamily{cmss}
\fontsize{17pt}{27pt}
\selectfont
\textbf{Darboux transformations and random point processes.}
\end{Large}\\
\bigskip
M. Bertola$^{\dagger\ddagger}$\footnote{bertola@mathstat.concordia.ca},  
M. Cafasso$^{\diamondsuit}$ \footnote{cafasso@math.univ-angers.fr}
\\
\bigskip
\begin{small}
$^{\dagger}$ {\em  Department of Mathematics and
Statistics, Concordia University\\ 1455 de Maisonneuve W., Montr\'eal, Qu\'ebec,
Canada H3G 1M8} \\
\smallskip
$^{\ddagger}$ {\em Centre de recherches math\'ematiques,
Universit\'e de Montr\'eal\\ C.~P.~6128, succ. centre ville, Montr\'eal,
Qu\'ebec, Canada H3C 3J7} \\
\smallskip
$^{\diamondsuit}$ {\em LUNAM Universit\'e, LAREMA, Universit\'e d'Angers\\ 2 Boulevard Lavoisier, 49045 Angers, France.}\\
\end{small}
\vspace{0.5cm}
{\bf Abstract} \end{center}

In this paper we describe a general method to derive formulas relating the gap probabilities of some classical determinantal random point processes (Airy, Pearcey and Hermite) with the gap probability of the same processes  with ``wanderers'', ``inliers'' and ``outliers''. In this way, we generalize the Painlev\'e--like formula found by Baik for the Baik--Ben Arous--P\'ech\'e distribution to many different cases, both in the one and multi--time case. The method is not ad-hoc and relies upon the notion of  Schlesinger transformations for Riemann--Hilbert problems.

\vspace{0.7cm}

\vskip 15pt

\tableofcontents

\section{Introduction}

The Airy process governs the universal fluctuations of the largest eigenvalue in several matrix models, the asymptotics of the longest increasing subsequences of random permutations \cite{BaikDeiftJohansson}, the last-passage time in directed percolation \cite{Johansson_Shape}, to mention a few models.
In the multi--time case, the Airy process originally appeared in the study of polynuclear growths models \cite{PrSp} and it is the main example of the probability distributions appearing in the KPZ universality class \cite{KPZ}, which attracted many attention both by physicists and mathematicians in the last years.\\
It has been proven in \cite{BaikBenArousPeche} for the one time case, and in \cite{BorPe,ADvM,AFvM} for the multi--time case, that through some properly chosen modifications of the probabilistic systems leading to the Airy process, one can obtain a whole family of different universality classes indexed by two sets of real parameters; for directed percolation, for instance, the modification consists in considering the case of the percolation in a lattice where some lines and/or columns are ``defective'', while in Dyson Brownian motions one has to consider the case of several particles starting and arriving to a given point plus some ``wanderers''. We refer to the cited works for more details. Concretely, in the one--time case, these new universality classes are described in terms of a kernel of the form
 \be\begin{array}{ll}
	K^{(\a,\b)}_{\Ai}(x,y):=  \ds\frac{1}{(2\pi i)^2} \int_{\gamma_L}\!\!\!dw\int_{\gamma_R}\!\!\!dz \frac{{\rm e}^{\frac{z^3}3 - \frac{w^3}3 - zx + wy}}{w-z}
	\frac {\prod(z-b_i)}{\prod(w-b_i)} \frac {\prod(w-a_j)}{\prod(z-a_i)},
\end{array}\label{introMTAirykernel}\ee
where $\a := \left\{a_1,\ldots,a_q\right\}$ and $\b := \{b_1,\ldots,b_m\}$ are the two sets of real parameters (see Fig. 1 below).\\
Given the celebrated Tracy--Widom result \cite{TracyWidomLevel} relating the Painlev\'e II equation to the Fredholm determinant of the Airy kernel, it is just natural to ask if analogue ``Painlev\'e--like'' formulas can be written for the Fredholm determinants of the kernels above, together with their multi--time versions. An answer for the one--time case and just one set of parameter has been given by Baik in \cite{BaikPformulas}. Already in his paper, Baik pointed out that his formula ((1.16) in loc. cit.) resembles the ones associated to Darboux transformations in the theory of integrable systems. 
We shall indeed explain this conjectured connection and also generalize it to the case of the Airy kernel with two sets of parameters, both in the one and in the multi--time case. In fact, our approach is rather general
and can be applied to a wide class of kernels (see the beginning of Section \ref{sectGP}); as additional examples motivated by the existing literature, we deduce some formulas for the gap probability of the Pearcey process with inliers (both in the one and multi--time case) \cite{ADvMVA} and for the gap probability associated to the Gaussian Hermitian matrix model with a finite--rank external potential \cite{BHExternalpotential,Johansson1,BleKui1}.\\

We now describe our approach in general terms: we consider a generic  Riemann--Hilbert problem for a $N\times N$ matrix $\Gamma(z;\un t)$ as follows
\be
\label{RHGamma}
\Gamma_+(z;\un t) = \Gamma_-(z;\un t) M(z;\un t)\ , z\in \Sigma; \ \ \ \Gamma(z;\un t) = \1 + \mathcal O(z^{-1})  \  ,\ \ z\to \infty.
\ee
where  $\un t$ denotes a set of deformation parameters on which the {\em jump matrix} $M(z;\un t)$ depends smoothly.
Let $\pa$ denote any derivative in one of the parameters $\un t$; here $\Sigma $ is a finite union of piecewise smooth, oriented  curves, and the subscripts $_\pm$ denote the (nontangential) boundary values from the left/right.
Let $\omega (\pa,[M] )$ denote the following differential on the manifold of parameters $\un t$
\be
\label{omegadiff}
\omega(\pa, [M]):= \int_{\Sigma } \frac {\d z}{2i\pi} \tr \bigg(\Gamma_-^{-1} \Gamma_-' \pa M M^{-1}.
\bigg)
\ee
The differential \eqref{omegadiff} was introduced in \cite{BertolaIsoTau}\footnote{The sign in loc. cit. is the opposite; however that is a ``mistake'' in the sense that the definition should have been posed as it is now. } as a generalization of the isomonodromic tau function invented by the Japanese school \cite{JMU1}; it  is known that the exterior differential $d \omega$ on two (commuting) derivatives $\pa, \wt \pa$ is 
\be
\d \omega (\pa, \wt \pa;[M]) =   \int_{\Sigma } \frac {\d z}{2i\pi} \tr \bigg(\frac {\d}{\d z}\pa M M^{-1}  \wt \pa M M^{-1}   -\pa M M^{-1} \frac {\d}{\d z} \wt \pa M M^{-1}  \bigg)
\ee
Another related differential was also introduced in \cite{BertolaIsoTau} as 
\bea
\Omega(\pa;[M]) := \frac 1 2  \int_{\Sigma } \frac {\d z}{2i\pi} \tr \bigg(\Gamma_-^{-1} \Gamma_-' \pa M M^{-1}  + 
\Gamma_+^{-1} \Gamma_+'  M^{-1} \pa M\bigg);\\
\d \Omega(\pa, \wt \pa;[M] ) = \frac 1 2  \int_{\Sigma } \frac {\d z}{2i\pi} \tr \bigg(M^{-1}M' \le[ M^{-1} \pa M, M^{-1} \wt \pa M^{-1}\ri]\bigg),
\eea
and the relationship between the two was established to be 
\be
\label{Omegaomega}
\Omega - \omega  = \frac 1 2 \int_{\Sigma } \frac {\d z}{2i\pi} \tr \bigg(M' M^{-1}  \pa M M^{-1}.
\bigg)
\ee

In many specific situations of interest, the differential $\Omega$ (or $\omega$, or both)  are  {\em closed} differentials and thus they locally define  a function on the space of parameters; 
in the case of {\em isomonodromic deformation} \cite{BertolaIsoTau} the precise relation is 
\be
\pa \ln \tau_{JMU}  = \Omega(\pa,[M])
\ee
where $\tau_{JMU}$ denotes the Jimbo-Miwa-Ueno isomonodromic tau function \cite{JMU1}, and $\pa$ here is an {\em isomonodromic} deformation. The identification of the differential of the tau--function with the differential $\Omega$ allows to introduce also formul\ae\ for the differentiation also  with respect {\em monodromy} parameters (e.g. the Stokes' parameters); we refer to \cite {BertolaIsoTau} for details.

In the case of gap probabilities of determinantal random processes with kernels as \eqref{introMTAirykernel} or, more generally, \eqref{generickernelMT} \cite{BertolaCafasso1,BertolaCafasso3}, the same differential \eqref{omegadiff} is directly related to the logarithmic derivative of a Fredholm determinant: the typical example is the Fredholm determinant defining the Tracy-Widom distribution \cite{TracyWidomLevel} in terms of the Airy kernel. 
In this context, our first result of a general nature addresses the following question:

\begin{question}
\label{ques1} 
{Given a Riemann--Hilbert problem as \eqref{RHGamma} with jump matrix $M(z,\t)$, how do $\omega(\pa; [M])$ and  $\Omega(\pa;[M])$ change under the group--action of conjugations of  the jumps by a diagonal rational matrix $D(z)$?}  
\end{question} 
The question is particularly relevant in the cases where $\omega$ is the differential of a function, be it an isomonodromic tau function or a Fredholm determinant according to the case under scrutiny: indeed in those cases the answer to Question \ref{ques1} implies a formula for the {\em ratio} of the tau-function/Fredholm determinant before and after the modification of the jumps. Following the literature, the transformation of the Riemann--Hilbert problem under the action of the diagonal matrix $D$ shall be referred as a \emph{Schlesinger transformation}, while we will speak about \emph{Darboux transformation} for the action on the corresponding tau functions.
The problem of finding the variation of the isomonodromic tau functions under the action of Darboux transformation had been addressed (and also solved) in \cite{JMU2,JMU3}, but in the narrower context of matrix differential equations. We will reformulate the problem and give a more constructive proof (not by induction) in the first part of the paper. 
The paper is organized as follows:
\begin{enumerate}
	\item In Section \ref{sectGT}, we formulate the problem of the variation of the isomonodromic tau function under Darboux transformation in the general setting of the Riemann--Hilbert problem and answer Question \ref{ques1} in full; the main Theorem is \ref{maintheorem}, whose proof will be given in the Appendix \ref{appMarco}.
	\item In Section \ref{sectGP}, we shall apply the general theory of the previous section to the so--called Baik--Ben Arous--P\'ech\'e distribution, its generalization to two sets of parameters and the multi--time case, the Pearcey process with inliers and the Hermite kernel with external potential; all these cases are specialization of Cor. \ref{corFratio}. In this section, we will also briefly describe some connections with the theory of KP equations. In particular, using the KP equation, we re--derive a PDE for the gap probability of the Airy process with two sets of parameters (in the one time case); a PDE that was originally found in \cite{ACvM1}, but for a slightly less general case.  
	\item Appendix \ref{AP} contains the Theorem \ref{thdetMalgrange}, proven in \cite{BertolaCafasso1} and used to identify, along the paper, the Fredholm determinants we are interested in with the isomonodromic tau function of some Riemann--Hilbert problems.
\end{enumerate}

\section{Riemann--Hilbert problems and Darboux transformations}\label{sectGT}
Let $\Sigma \subset \C$ be a collection of piecewise smooth oriented arcs and let 
us consider a Riemann--Hilbert problem for the $N\times N$ matrix--valued,  piecewise analytic  and analitycally invertible function  $\Gamma(z)$ (depending parametrically on $\un t$) 
\be\label{RHPGamma}\left\{\begin{array}{ccc}
		 \Gamma_+(z,\un t)&=& \Gamma_-(z,\un t) M(z, \un t),\quad z\in\Sigma,\\
		\\
		 \Gamma(z, \un t)& =& \Id+\mathcal O(z^{-1}),\quad z \longrightarrow\infty.
	\end{array}\right.
\ee
We shall not enter into analytic details on specific requirements on $M$ at this stage; we tacitly assume that the RHP \eqref{RHPGamma} is {\em generically} (in the parameter space $\un t$) solvable. 
Let 
$D(z)$ be a diagonal rational matrix. We will denote with $\mathcal B$ the set of poles of $D(z)$ (by which we mean a pole of any of the entries) and with $\mathcal A$ the set of poles of $D^{-1}$ (i.e. zeroes of $D(z)$). 
We shall denote the multiplicity  of the pole  $z=b\in \mathcal B$ in the $\mu$-th entry by $k_{b,\mu}$ and similarly the order of the pole $z=a\in \mathcal A$ in the $\nu$-th entry by $\ell_{a,\nu}$ and by $K_b = {\rm diag}(k_{b,1}, \dots , k_{b, N}),\ L_a = {\rm diag}(\ell_{a,1}, \dots , \ell_{a, N})$ so that 
\be
D(z) = \prod_{c\mathcal A\cup \mathcal B} z_c^{L_c-K_c} 
\ee
where $L_c=0$ if $c \not\in \mathcal A $ and $K_c = 0$ if $c\not \in \mathcal B$. 
\bd
\label{defGmatrix}
The {\bf characteristic matrix}\footnote{We borrow the name from \cite{JMU2}.} $G = G_{
{\le\{{\mathcal A\, \mathcal B}\atop{ L\,K}\ri\}}
}$ associated to the Riemann--Hilbert problem \ref{RHPGamma} and the diagonal matrix $D$  is the following matrix:  
\bea
\label {g13}
G_{(b,\nu,k); (a,\mu,\ell)} :=  \res_{z=a}\res_{\z=b}\frac { {\bf e}_\mu^t \Gamma^{-1}(z) \Gamma(\z) {\bf e}_\nu   \d z \d \z}{(z_a)^{\ell- \delta_{a\infty}} (z-\z) (\z_b)^{k_{b,\nu}+1-k  - \delta_{b\infty}}},
\eea
where $z_{c} := (z-c)$ if $c$ is a finite point and $z_\infty := 1/z$ if $c=\infty$, and where $b$ runs amongst all the poles of $D(z)$, $a$ amongst the poles of $D^{-1}(z)$ and $1 \leq k\leq  k_{b,\nu}$, $1 \leq \ell \leq  \ell_{a,\mu}$, $1\leq \mu,\nu\leq N$. This matrix has size $\sum_{b\in \mathcal B}  \sum_{\nu=1}^N k_{b, \nu}$  (note that the total number of poles/zeroes is equal, counting the poles /zeroes at infinity).
\ed
\br
Since $D(z)$ cannot have both a pole and a zero in the same entry at the same point $c$, then we never have $(b,\nu) = (a,\mu)$ and then the reader may verify that the order of the residues in \eqref{g13} is  immaterial.
\er
\begin{example}
Simple examples are those where $\mathcal A, \mathcal B$ are singletons and  the  $L, K$ are elementary diagonal matrices $E_{\mu\mu}$ so that the characteristic matrix is $1\times 1$. This corresponds to the case of ``elementary'' Schlesinger transformations\footnote{For the definition of Schlesinger transformation we used see Definition \ref{defSchles}} of \cite{JMU2} (see Sec. \ref{secSchles});
\be
\begin{array}{ll}
G_{\le\{\{c\}\ \{c\}\atop E_{\mu\mu}\ E_{\nu\nu}\ri\}} = (\Gamma^{-1}(c)\Gamma'(c))_{\mu\nu}  \ (\mu\neq \nu);&
G_{\le\{\{\infty\}\ \{b\}\atop E_{\mu\mu} \ E_{\nu\nu} \ri\}} = (\Gamma(b))_{\mu\nu}   ,\ b\neq \infty\\[14pt]
G_{ \le\{\{a\}  \ \{\infty\}\atop E_{\mu\mu}\ E_{\nu\nu} \ri\} } = (\Gamma(a)^{-1})_{\mu\nu} ,\ a \neq\infty; & 
\ds G_{\le\{\{a\}\  \{b\} \atop E_{\mu\mu}\ E_{\nu\nu} \ri\}}= \frac{(\Gamma^{-1}(a)\Gamma(b))_{\mu\nu}}{a-b}    \ , a\neq \infty\neq b.
\end{array}
\label{taudets}
\ee
\end{example}
Our first main theorem characterizes the change that the isomonodromic differential $\omega(\pa,[M])$ \eqref{omegadiff} undergoes under a conjugation of the jump matrix by the diagonal matrix $D(z)$. First of all we have the theorem that explains how the characteristic matrix is the coefficient matrix of a linear problem.
\bt
\label{maintheorem1}
Let $\Gamma$ be the solution of the Riemann--Hilbert problem \ref{RHPGamma} and $D$ a diagonal {\bf rational} matrix. Assume moreover that none of the poles and zeroes of $D$ lies on $\Sigma$\footnote{See Remark \ref{formalrem} for a relaxation of this assumption when possible. The statement of the theorem is unchanged.}.
Let  $R(z)$ be a  rational matrix with poles at $\mathcal A\cup \mathcal B$ and  such that\\
{\bf [a]} $\det R(z) \neq 0, \ \ z\not\in \mathcal A\cup \mathcal B$,\\
{\bf [b]}  near each pole and zero of $D(z)$, the new matrix $\wt \Gamma(z;\un t):= R(z;\un t) \Gamma(z;\un t) D(z)$ is regular, namely 
\be
(R(z) \Gamma(z) D(z))^{\pm 1}  = \mathcal O(1)\ , z\to c\ \ \
R (z)\Gamma(z) D(z) = \1 + \mathcal O(z^{-1}) , \ \ z\to\infty,
\ee
where $c$ denotes a pole of $D(z)$ or $D^{-1}(z)$, respectively.\\
 Then the matrix $R(z)$ exists if and only if $\det G_{
 {\le\{{\mathcal A\, \mathcal B}\atop{ L\,K}\ri\}}}\neq 0$ (with $G_{{\le\{{\mathcal A\, \mathcal B}\atop{ L\,K}\ri\}}}$ the characteristic matrix related to $\Gamma$ as in Def. \ref{defGmatrix})
 \et
 The proof is  App. \ref{proofthm1}.
 Secondly we shall prove 
 
 \bt
 \label{maintheorem}
  Let $\wt \Gamma(z;\un t):= R(z;\un t) \Gamma(z;\un t) D(z)$ as in Thm. \ref{maintheorem1}. Then:\\ 
 {\bf [1]} $\wt \Gamma$ solves the following new Riemann--Hilbert problem 
\be
\wt \Gamma(z;\un t, \un c)_+ = \wt \Gamma(z;\un t, \un c)_- \wt M(z; \un t, \un c)\ ,\qquad    \wt \Gamma(z) = \1 + \mathcal O(z^{-1})\ ,\ z\to\infty,  \ \det \wt \Gamma (z)\neq 0\ ,\ \ \forall z.
\ee
where $\un c$ denotes symbolically the acquired dependence on the poles/zeroes of $D(z)$  and the new jump matrix is given by 
\be
\wt M(z; \un t, \un c) := D^{-1}(z;\un c) M(z;\un t) D(z;\un c)\ ,
\label{conjD}
\ee
{\bf [2]} The difference of the two isomonodromic differentials \eqref{omegadiff} is 
\bea
\omega(\pa, [\wt M]) - \omega(\pa, [M]) =&\&  \sum_{c} \left(\res_{z=c} \tr \le(
R^{-1} R' \pa (\Gamma D) D^{-1} \Gamma^{-1} 
\ri) +
\res_{z=c} \tr \le(
\Gamma^{-1}\Gamma' \pa D D^{-1}\ri)\right)+\nonumber\\
&\&  +  \int_\Sigma\! \frac{dz}{2\pi i} \tr \le( D^{-1} D' \le(  
\pa M M^{-1}  - \pa D D^{-1} + MD^{-1} \pa D M^{-1} \ri) - M^{-1} M' \pa D D^{-1}\ri),
\eea
where $\pa$ denotes now a derivative of any of the  parameters $\un t$ or $\un c$. \\
{\bf [3]}
Equivalently, 
\bea
\label{omegaSchles}
\omega(\pa, [\wt M]) - \omega(\pa, [M]) =
\pa \ln \det G
+ \pa \ln \prod_{\mu=1}^n
\frac {\prod_{b\in \mathcal B' \atop a \in \mathcal A'} (b-a)^{k_{b,\mu} \ell_{a,\mu}}} 
{ \prod_{b<b'} (b-b')^{k_{b,\mu}k_{b',\mu}}\prod_{a<a'} (a-a')^{\ell_{a,\mu} \ell_{a',\mu}}} 
 +\nonumber\\
 +  \int_\Sigma \frac{dz}{2\pi i} \tr \le( D^{-1} D' \le(  
\pa M M^{-1}  - \pa D D^{-1} + MD^{-1} \pa D M^{-1} \ri) - M^{-1} M' \pa D D^{-1} \ri),
 \eea
with $G$ the relevant characteristic matrix  in Definition \ref{defGmatrix}, and $b<b'$ denotes any (fixed) ordering of the {\em finite} points in $\mathcal B' := \mathcal B\setminus \{\infty\}$ ($\mathcal A' := \mathcal A \setminus \{\infty\}$).
\et
A simple corollary expressing the change of $\Omega$ (rather than $\omega$) is given below; 
 \bc
 Under the same hypothesis of Theorem \ref{maintheorem} we have (all products over $b\in \mathcal B' = \mathcal B\setminus \{\infty\}$ and $a\in \mathcal A' = \mathcal A/\{\infty\}$)
 \bea
 \label{117}
\Omega(\pa, [\wt M]) - \Omega(\pa, [M]) =
\pa \ln \det G
+ \pa \ln \prod_{\mu=1}^n
\frac {\prod_{b\in \mathcal B' \atop a \in \mathcal A'} (b-a)^{k_{b,\mu}\ell_{a,\mu}}} 
{ \prod_{b<b'} (b-b')^{k_{b,\mu}k_{b',\mu}}\prod_{a<a'} (a-a')^{\ell_{a,\mu}\ell_{a',\mu}}} 
 +\nonumber\\
 \label{118}
  +  \frac 12 \int_\Sigma \frac{\d z}{2i\pi} \tr \bigg(
  D' D^{-1} \le( \pa M M^{-1}+ M^{-1} \pa M  + M \pa D D^{-1}M^{-1} -M^{-1} \pa D D^{-1} M \ri) +\nonumber\\
   \label{119}
  -    \pa D D^{-1} \le(M^{-1}M' + M'M^{-1}\ri)\bigg)
\eea
\ec
{\bf Proof}. 
Using the relationship \eqref{Omegaomega} we only have to add to \eqref{omegaSchles} in Thm. \ref{maintheorem}   the expression  below
\bea
\frac 1 2 \int_\Sigma \frac {\d z}{2i\pi} 
\tr \bigg(\wt M' \wt M^{-1} \pa \wt M \wt M^{-1}  -  M'  M^{-1} \pa  M  M^{-1} \bigg),
\eea
with $\wt M  = D^{-1}MD$. Simplifying the result (using the cyclicity of the trace only) we obtain the claim. 
{\bf Q.E.D.}

The Theorems \ref{maintheorem1} and particularly \ref{maintheorem} are quite crucial and actually costitute the largest conceptual contribution of the paper. However we opted to confine their proofs to the Appendix (Section \ref{proofmain}) in order not to interfere with the momentum of the paper. In the setup of this theorem $D(z)$ is a globally defined rational function, therefore each entry has total index zero on the Riemann sphere (the index is the difference between the total multiplicity of zeroes and poles, including infinity). In fact we will prove first a more general statement where $D(z)$ is only locally defined in disjoint disks, so that the entries may have total index different from zero. We shall anyway always assume that the index of the $\det D(z)$ (i.e. the sum of all the multiplicities of all the zeroes and all the poles of the entries) is zero (in Section \ref{secSchles}). 
\section{Applications to random point processes}\label{sectGP}
We will only consider gap probabilities involving at most two-times because it offer a sufficient generality without unnecessary technical complications; the completely general formulas would only require writing larger matrices and  was considered in \cite {BertolaCafasso3}.  
We  thus consider the evaluation of the Fredholm determinant
\bea\label{genericFmt}
	F^{(\a,\b)}(\vtau,\s) &:=& \det(\Id - K^{(\a,\b)}\chi_{\I}) \\
	&:=& 1 + \sum_{k = 1}^\infty\sum_{r_1 + r_2 = k} \ds\int_{I_1^{r_1}\times I_2^{r_2}}\!\!\!\! \det\Bigg(\left(K_{ij}^{(\a,\b)}(x^{(i)}_s,x^{(j)}_t)\right)_{s \leq r_i, t \leq r_j}\Bigg)_{i,j = 1}^2 \prod_{i = 1}^{r_1}dx_i^{(1)} \prod_{i = 1}^{r_2}dx_i^{(2)}, \nonumber
\eea 
where we have two time parameters $\vtau := \{\tau_1,\tau_2\}$,  $\tau_1 < \tau_2$; $\I := \{I_1,I_2\}$ is a collection of two multi--intervals\footnote{By "multi-intervals" we always mean, following the general usage, a finite union of disjoint intervals.} with endpoints $\s = \{s_i^{(j)}, j =1,2\}$. The sets $\mathcal A' := \{a_j, j =1,\ldots, m_1\},\, \mathcal B' := \{b_i, i =1,\ldots m_2 \}$ consist of real parameters; to each $b_i$ we assign multiplicity $k_i$ and to each $a_j$ multiplicity $\ell_j$. The operator $K^{(\a,\b)}:L^2(\R, \C^2) \to L^2(\R,\C^2)$ is thought of as a matrix integral operator with a smooth kernel (denoted by the same symbol for simplicity) and entries given by
\be
	K^{(\a,\b)}_{ij}(x,y) := \frac{1}{(2\pi i)^2} \int_{\gamma_2}\!\!\!\!dw\!\!\int_{\gamma_1}\!\!\!\!dz \frac{{\rm e}^{\theta_{\tau_i}(x;z)-\theta_{\tau_j}(y;w)}}{w-z}
	\frac {C(z)}{C(w)}
 -\delta_{i < j} \!\!\int _{\gamma_2}\!\! \frac{d z }{2\pi i} {\rm e}^{\theta_{\tau_i}(x; z )-\theta_{\tau_j}(y; z )}.\nonumber\\\label{generickernelMT}
\ee
where $C(z):= \ds\frac{\prod_{j=1}^{m_1} z_{b_j}^{k_j}}{\prod_{i = 1}^{m_2} z_{a_i}^{\ell_j}}$ (same notations as in the previous section).

\paragraph{Overview of the logic.}
The logical development of this section and all the specializations is as follows: 
\begin{itemize}
\item  We associate the  RHP  \eqref{genericRH1MT} to the operator in \eqref{genericFmt}, for arbitrary number of $\a,\b$'s. This RHP is of the general form \eqref{RHPGamma}. 
\item The jump matrices $M^{(\a,\b)}(z; \vtau, \s)$ of the RHP depend on the parameters $\vtau, \s$ and $\a,\b$; the dependence on these parameters of the Fredholm determinant $F^{(\a,\b)}(\vtau,\s)$ \eqref{genericFmt} is exactly an instance of isomonodromic differential \eqref{omegadiff}. This means that 
\be
\pa \ln F^{(\a,\b)}(\vtau,\s)  =  \omega (\pa; [M^{(\a,\b)}])
\ee
\item The jump matrices $M^{(\a,\b)}$ for different number of points $\a,\b$'s are all related to the ``bare'' jump matrices $M^{(\emptyset ,\emptyset)}$  by a conjugation with a diagonal rational matrix; therefore we are precisely in the setting of Thm. \ref{maintheorem}. 
\item As a consequence, we shall  establish (Cor. \ref{corFratio}) that 
\be
\frac {F^{(\a,\b)}(\vtau,\s)}{F^{(\emptyset, \emptyset)}(\vtau,\s)} \propto \frac {\ds \det G_{\le\{{\mathcal A\, \mathcal B}\atop{ L\,K}\ri\}}
{\ds \prod_{j\leq m_1, r \leq m_2} (b_j-a_r)^{k_j \ell_r}}}  
{ \ds \prod_{1\leq j< j'\leq m_1} (b_{j}-b_{j'})^{k_{j}k_{j'}}\prod_{1\leq r<r'\leq m_2} (a_{r}-a_{r'})^{\ell_{r} \ell_{r'}}}  
\label{Fratio}
\ee
where the proportionality constant is an absolute constant (independent of $\vec a, \vec b, \vtau,\s$) whose specific evaluation depends on the example and will then be shown to be unity.
\end{itemize}

We make the following assumptions, which cover all the cases to be detailed below:
\begin{enumerate}
	\item $\theta_\tau(x;z)$ is a polynomial in $z$ whose coefficients (possibly) depend on a real parameter $\tau$. Moreover $\theta_\tau(x;z)$ is linear in $x$; more precisely $\theta_\tau(x;z) = \theta_\tau(0;z) - xz$.
	\item $\gamma_2$ can be continuously deformed into $i\R$ without changing the value of the kernel and $\gamma_1$ is (union of) oriented contours  (possibly extending to infinity)  such that the integrand on the left hand side of \eqref{generickernelMT} is convergent and $(\gamma_1 \cap \gamma_2) \cap \{\mathcal A' \cup\mathcal B'\} =\emptyset$. Of course, also the series in \eqref{genericFmt} is assumed to be convergent.
	\item $\gamma_1$ has no intersections with $\gamma_2$. If $\mathrm{Re}(z - w) > 0$ for all $z \in \gamma_1, w\in \gamma_2$, then an odd number of endpoints for $I$ is allowed and $I = [s_1,s_2] \sqcup \ldots \sqcup [s_{2k+1},+\infty)$. Viceversa, if  $\mathrm{Re}(z - w) < 0$ for all $z \in \gamma_1, w \in \gamma_2$, then $I = (-\infty, s_1] \sqcup [s_2,s_3] \sqcup \ldots \sqcup [s_{2k},s_{2k+1}]$. In the other cases just an even number of endpoints is considered.
\end{enumerate}

\begin{remark}\label{wrongsign}
	In concrete examples coming from the literature the kernel is as in \eqref{generickernelMT} but with  $\wt\theta_\tau(x;z) = \wt\theta_\tau(0;z) + xz$ (thus differing in the sign of $x$). The change is only cosmetic as one promptly verifies that it only amounts to a reversal of sign of all the multi-intervals.
\end{remark}
 Throughout the section we shall pose $N = N_1 + N_2 = \# s^{(1)} + \# s^{(2)}$.
 We introduce the following RHP:
\begin{problem}\label{genericRH1MT}
Find the sectionally analytic function $\Gamma^{(\a,\b)}( z ) \in \mathrm{GL}(N+1,\C)$ on $\C/\left\{\gamma_1 \cup \gamma_2\right\}$ such that
\be\label{genericRHeqMT}
	\left\{\begin{array}{ll}
	\Gamma_+^{(\a,\b)}( z ) =  \Gamma_-^{(\a,\b)}( z )M^{(\a,\b)}( z )&\\
	\\
	\Gamma^{(\a,\b)}( z )  \sim  \1 + \mathcal O( z ^{-1}), \quad  z  \rightarrow \infty .&
	\end{array}\right.					 
\ee
with\footnote{In order to make the formula readable, in the expression of the jump matrix we suppressed the dependence of the variables on $ z $.} $M^{(\a,\b)}( z ) :=$
\begin{small}\be
	\nonumber \left[\begin{array}{c|c|c}
		1 & -{\rm e}^{\theta_{\tau_1}\left(s_1^{(1)}\right)}C\chi_1 \  \ldots \  (-)^{N_1}{\rm e}^{\theta_{\tau_1}\left(s_{N_1}^{(1)}\right)}C\chi_1 & -{\rm e}^{\theta_{\tau_2}\left(s_1^{(2)}\right)}C\chi_1 \  \ldots \  (-)^{N_2}{\rm e}^{\theta_{\tau_2}\left(s_{N_2}^{(2)}\right)}C\chi_1\\ \hline
	\rule{-1pt}{15pt} -{\rm e}^{-\theta_{\tau_1}\left(s_1^{(1)}\right)}C^{-1}\chi_2 &   & \\
	 \vdots &   \1_{N_1}   & \bf{0}  \\ 
	  -{\rm e}^{-\theta_{\tau_1}\left(s_{N_1}^{(1)}\right)}C^{-1}\chi_2 &    & \\
	   \hline
	  \rule{-1pt}{15pt}-{\rm e}^{-\theta_{\tau_2}\left(s_1^{(2)}\right)}C^{-1}\chi_2 &   & \\
	 \vdots &   R_{ij} &   \1_{N_2} \\ 
	  -{\rm e}^{-\theta_{\tau_2}\left(s_{N_2}^{(2)}\right)}C^{-1}\chi_2   &  & 
	\end{array}\right]
\ee \nonumber\\
\end{small}
$$R_{ij}( z ) := (-1)^{j+1}{\rm e}^{-\theta_{\tau_1}(s_j^{(1)}; z )+\theta_{\tau_2}(s_i^{(2)}; z )}\chi_2,\quad i = 1,\ldots, N_2,\,\, j = 1, \ldots , N_1.$$
Here $\chi_{1,2} = \chi_{1,2}(z)$ denote the indicator functions of $\gamma_{1,2}$ and $C = C(z) = C(z; \vec a, \vec b)$ is given by 
\be
\label{defC}
C(z):= \ds\frac{\prod_{j=1}^{m_1} (z-{b_j})^{k_j}}{\prod_{i = 1}^{m_2} (z-{a_i})^{\ell_j}}.
\ee
\end{problem}
\begin{theorem}\label{genericThmMT}
The Fredholm determinant $F^{(\a,\b)}$ satisfies the following variational formul\ae
\be
\pa \ln F^{(\a,\b)}(\vtau, \s) = \omega(\pa; [M^{(\a,\b)}])
\ee
with $\omega$ as in \eqref{omegadiff}.
In particular, 
\be\label{genericlogderivativesMT}
		\partial_{s_i^{(1)}} \log F^{(\a,\b)} = -\left(\Gamma_1^{(\a,\b)}\right)_{(i+1,i+1)},\quad \partial_{s_i^{(2)}} \log F^{(\a,\b)} = -\left(\Gamma_1^{(\a,\b)}\right)_{(N_1+i+1,N_1+i+1)}
	\ee
	where
	\be\label{genericasympexpansionMT}
		\Gamma^{(\a,\b)}\sim \1 + \Gamma_1^{(\a,\b)} z ^{-1} + \mathcal{O}( z ^{-2}),\quad  z \rightarrow\infty.
	\ee
\end{theorem}
{\bf Proof.}
The proof is essentially a rewrite of the proof of Thm.  2.1 of \cite{BertolaCafasso3}, and we report it here mostly for the convenience of the reader.
We start observing that, using the Cauchy residue's theorem on the  $(i,j)$ entry of the kernel, we have
	\bea
	&&K_{i,j}^{(\a,\b)}(x,y)\chi_{\I}(x)  =  \frac{1}{(2\pi i)^3} \int_{i\bR}\!\!\! d\xi \sum_{\ell = 1}^{N_i}(-1)^{\ell + 1}{\rm e}^{\xi(s_\ell^{(i)}-x)} \nonumber \\
	&&\times \left[ \int_{\gamma_2}\!\!\! dw\int_{\gamma_1}\!\!\! dz \frac{{\rm e}^{\theta_{\tau_i}(s_\ell^{(i)};z)-\theta_{\tau_j}(0;w)+yw}}{(z-w)(\xi-z)}
	\frac{C(z)}{C(w)} +
	  2\pi i  \delta_{1i}\delta_{2j} \int_{\gamma_2} dz {\rm e}^{\theta_{\tau_i}(s_{\ell}^{(i)}:z)-\theta_{\tau_j}(0;z)+yz} \right].
	\nonumber \eea
Hence, in particular, $K^{(\a,\b)} = \mathcal{T}K^{(\a,\b;\s)}\mathcal{T}^{-1},$ where $\mathcal{T}: L^2(\gamma_2)\longrightarrow L^2(\R)$ is the Fourier transform
and the entries of $K^{(\a,\b;\s)}$ are given by
\bea
	K_{ij}^{(\a,\b;\s)}(\xi,w) :=\sum_{\ell = 1}^{N_i}(-1)^{\ell + 1}\left[\int_{\gamma_1}\!\! \frac{dz}{2\pi i} \frac{{\rm e}^{\theta_{\tau_i}(s_\ell^{(i)};z)-\theta_{\tau_j}(0;w)+s_{\ell}^{(i)}\xi}}{(z-w)(\xi-z)}
	\frac{C(z)}{C(w)}\!\! + \delta_{1i}\delta_{2j}\frac{{\rm e}^{\theta_{\tau_i}(s_{\ell}^{(i)};w)-\theta_{\tau_j}(0;w)+s_\ell^{(i)}\xi}}{\xi-w}\right].\nonumber
\eea
$K^{(\a,\b;\s)}$ is the composition of two integrable operators plus a third one, namely
\be
\label{KGFE}
K^{(\a,\b;\s)}(\xi,w) = \mathcal G_\s(\xi,z)\ast \F(z,w) + \mathcal{E}(\xi,w)
\ee
with
\be\nonumber
	\begin{array}{cc}
		(\mathcal G_\s)_{ij}(\xi,z) := \ds\frac{\delta_{ij}}{2\pi i}\ds\sum_{\ell = 1}^N(-1)^{\ell + 1}\ds \ds\frac{{\rm e}^{\frac{1}2\theta_{\tau_i}(0;z)-s_\ell^{(i)}(z-\xi)}}{\xi-z}; & \mathcal G_\s : L^2(\gamma_1) \longrightarrow L^2(\gamma_2),\\
		\F_{ij}(z,w) :=  \ds\frac{1}{2\pi i} \ds\frac{{\rm e}^{\frac{1}2\theta_{\tau_i}(0;z)-\theta_{\tau_j}(0,w)}}{z - w}
		\frac{C(z)}{C(w)};& \F : L^2(\gamma_2)\longrightarrow L^2(\gamma_1),\\
		\mathcal{E}_{ij}(\xi,w) := \delta_{1i}\delta_{2j}\ds\sum_{\ell = 1}^{N_1}\frac{{\rm e}^{\theta_{\tau_1}(s_\ell^{(1)};w)-\theta_{\tau_2}(0;w)+s_{\ell}^{(1)}\xi}}{\xi - w} ; & \mathcal{E} : L^2(\gamma_2) \longrightarrow L^2(\gamma_2).
	\end{array}
\ee
Now define the following integral operator acting on $L^2(\gamma_1\sqcup\gamma_2) = L^2(\gamma_1) \oplus L^2(\gamma_2)$:
$$ H :=\left[\begin{array}{cc}
						0 & \F\\
						\mathcal G_\s & \mathcal{E}
					\end{array}\right]. $$
The operator $H$ is easily seen to be HS and thus $\det_2(\Id-H)$ is well defined, so that we have the following chain of identities of regularized Fredholm determinants \cite{Si}
\bea\nonumber
{\det}_2 (\Id -H) &\&= {\det}_2\left[\begin{array}{cc}
					\Id & -\F\\
					-\mathcal G_\s & \Id - \mathcal{E}
			\end{array}\right] = 
{\det}_2 \left[\begin{array}{cc}
	\Id & -\F\\
	-\mathcal G_\s & \Id - \mathcal{E}
\end{array}\right]{\det}_2 \left[\begin{array}{cc}	\Id & \F\\
 0 & \Id
\end{array}\right] = \\
\nonumber\\ \nonumber
 &\&={\rm e}^{-\tr\mathcal G_\s\ast\F} {\det}_2\left[\begin{array}{cc}
					\Id & 0\\
					-\mathcal G_\s & \Id -\mathcal G_\s\ast\F - \mathcal{E}
			\end{array}\right]
			={\rm e}^{-\tr\mathcal G_\s\ast\F}  {\det}_2(\Id - K^{(\a,\b:\s)}) \label{312}
\eea
The term $\tr\mathcal G_\s\ast\F$ can be regarded as the formal ``trace'' of $K^{(\a,\b:\s)}$ in \eqref{KGFE} because the term $\mathcal G_\s\ast\F$ is honestly of trace class (both operators are easily shown to be HS), but it is not obviuos whether $\mathcal E$  is. However $\mathcal E$ is diagonal free and hence (formally) traceless. 
Now the operator $K^{(\a,\b:\s)}$ is easily seen to be at least Hilbert--Schmidt and thus its $\det_2$ is well defined. The expression \eqref{312} then (by the properties of the regularized determinant $\det_2$) is computed in terms of the same series that defines an ordinary Fredholm determinant. On the other hand, since the operator $K^{(\a,\b:\s)}$ is unitarily equivalent to $K^{(\a,\b)}\chi_{I}$ by Fourier transform, the latter determinant equals the Fredholm determinant of $K^{(\a,\b)}\chi_{I}$. The reader interested in more details of the functional analysis behind this step is invited to refer to Thm. 2.1 of \cite{BertolaCafasso3}. 
Thus we have
\be\nonumber
		F^{(\a,\b)}  = {\det}_2(\Id - H).
\ee
It is then straightforward, although it takes some effort, to see   that the kernel of $H$ (which we denote by the same symbol) is $H(z,w)=\frac{f^{\mathrm{T}}( z )g(w)}{z-w}$ where $f$ and $g$ are the following matrices:
\be\nonumber
	f( z ) := \frac{1}{2\pi i} \left(\begin{array}{ccc} 
	{\rm e}^{\frac{1}{2}\theta_{\tau_1}(0; z )}C\chi_1 &
 	{\rm e}^{\frac{1}{2}\theta_{\tau_2}(0; z )}C\chi_1
	\\
	{\rm e}^{s_1^{(1)} z }\chi_2 & 0\\
	\vdots  & \vdots \\
	{\rm e}^{s_{N_1}^{(1)} z }\chi_2 & 0
	\\
	0 & {\rm e}^{s_1^{(2)} z }\chi_2\\
	\vdots & \vdots 
	\\
	0 &  {\rm e}^{s_{N_2}^{(2)} z }\chi_2
	\end{array}\right),\\
	\nonumber\\
	\nonumber
	g(w) :=  \left(\begin{array}{ccc} 
	{\rm e}^{-\theta_{\tau_1}(0;w)}C^{-1}\chi_2 & {\rm e}^{-\theta_{\tau_2}(0;w)}C^{-1}\chi_2
	\\
	{\rm e}^{\frac{1}2\theta_{\tau_1}(0,w)-s_1^{(1)}w}\chi_1 & {\rm e}^{\theta_{\tau_1}(s_1^{(1)},w)-\theta_{\tau_2}(0;w)}\chi_2
	\\
	\vdots & \vdots
	\\
	(-1)^{N_1+1}{\rm e}^{\frac{1}2\theta_{\tau_1}(0,w)-s_{N_1}^{(1)}w}\chi_1 & (-1)^{N_1}{\rm e}^{\theta_{\tau_1}(s_{N_1}^{(1)},w)-\theta_{\tau_2}(0,w)}\chi_2
	\\
	0 & {\rm e}^{\frac{1}2\theta_{\tau_2}(0;w)-s_1^{(2)}w}\chi_1
	\\
	\vdots & \vdots
	\\
	0 & (-1)^{N_2+1}{\rm e}^{\frac{1}2\theta_{\tau_2}(0,w)-s_{N_2}^{(2)}w}\chi_1
\end{array}\right).
\ee
Hence, $H$ is an integrable operator. Direct computation following the general theory explained in App. \ref{AP} shows that it is  associated to the Riemann--Hilbert problem \ref{genericRH1MT}.
In order to  prove  \eqref{genericlogderivativesMT} we need to transform the  Riemann-Hilbert problem to one with constant jump; to this end we define the diagonal matrix $T(z)$ as 
\bea\label{matrixTMT}
	T( z ) &\&:= \mathrm{diag}\left(T^{(0)}( z ),T^{(1)}( z ),\ldots,T^{(N)}( z )\right) + \log(C(z))\mathrm{diag}(1,0,\ldots,0)\\
	&\&T^{(0)}( z ) := \nonumber
	\ds\frac{1}{N+1}\left(\sum_{\ell = 1}^{N_1} \theta_{\tau_1}(s_\ell^{(1)}; z ) +\sum_{\ell = 1}^{N_2} \theta_{\tau_2}(s_\ell^{(2)}; z )\right),\\
	&\&T^{(\ell)}( z ) := T^{(0)}( z ) - \theta_{\tau_1}(s_\ell^{(1)}; z ),\;\ell = 1,\ldots, N_1,
;\qquad 	T^{(N_1 + \ell)}( z ) = T^{(0)}( z ) - \theta_{\tau_2}(s_\ell^{(2)}; z ),\;\ell = 1,\ldots, N_2. \nonumber
\eea
Then the reader may verify directly that 
$\Psi^{(\a,\b)}( z ) := \Gamma^{(\a,\b)}( z ){\rm e}^{T( z )}$ satisfies a RH problem with constant jumps.
The equation \eqref{genericlogderivativesMT} is then proved using \eqref{appisomtau} and verifying the equations
\bea
	&-\ds\res_{ z  = \infty}\mathrm{Tr}\left(\left(\Gamma^{(\a,\b)}\right)^{-1}\partial_ z  \Gamma^{(\a,\b)}\partial_{s_\ell^{(1)}} T\right)& = -	(\Gamma_1)_{\ell+1,\ell+1},\quad \ell = 1,\ldots,N_1;\nonumber\\
	&-\ds\res_{ z  = \infty}\mathrm{Tr}\left(\left(\Gamma^{(\a,\b)}\right)^{-1}\partial_ z  \Gamma^{(\a,\b)}\partial_{s_\ell^{(2)}} T\right)& = -	(\Gamma_1)_{\ell+1,\ell+1},\quad \ell = N_1+1,\ldots,N_1+N_2.\nonumber
\eea
 \QED
\subsubsection*{Relationship with Darboux transformations.}
In the Riemann--Hilbert problem \ref{genericRH1MT}
the jump matrices of the problem with $\a,\b\neq \emptyset$ are related to the jump matrices of the ``bare'' problem with $\a = \b = \emptyset$ by a diagonal conjugation 
\be\label{standardconjugation}
M^{(\a,\b)} (z)= D^{-1}(z) M^{(\emptyset,\emptyset)} (z) D(z)\ ,\ \ 
D(z) := {\rm diag}(C^{-1} (z), 1, \dots, 1)\ ,\quad C(z):= \ds\frac{\prod_{j=1}^{m_1} z_{b_j}^{k_j}}{\prod_{i = 1}^{m_2} z_{a_i}^{\ell_j}}.
\ee
This is a special case of Theorem \ref{maintheorem} (and also of Remark \ref{formalrem} when $\#\mathcal A' \neq \#\mathcal B'$) and hence
we have the following corollary :
\bc
\label{corFratio}
The Fredholm determinants $F^{(\a,\b)}(\vtau, \s)$ for different number of points $\a,\b$'s, are related as follows:
\bea
\label{Darbouxdiff+-}
\pa \ln F^{(\a,\b)}(\vtau,\s)  =\pa  \ln F^{(\emptyset,\emptyset)} (\vtau, \s) + \pa \ln \det G_{{\le\{{\mathcal A\, \mathcal B}\atop{ L\,K}\ri\}}
} + \pa \ln
\frac {\prod_{j\leq m_1\atop r\leq m_2} (b_j-a_r)^{k_{j} \ell_{r}}} 
{ \ds\prod_{j<j'} (b_j-b_{j'})^{k_{j}k_{j'}}\prod_{r<r'} (a_r-a_{r'})^{\ell_{r} \ell_{r'}}} \ ,\ \
\eea
where $\mathcal A$ and $\mathcal B$ are respectively the set of zeroes and poles of $D$; $K_{b_j} = k_j E_{11}$ and $L_{a_r} = \ell_r E_{11}$ while $\pa$ stands for the derivative with respect any of the parameters $\a,\b,\vtau,\s$. Equivalently, there is a constant $C$, independent of $\a,\b,\vtau,\s$ such that
\be
\frac {F^{(\a,\b)}(\vtau,\s)}{F^{(\emptyset, \emptyset)}(\vtau,\s)}= C \frac {\ds \det G_{\le\{{\mathcal A\, \mathcal B}\atop{ L\,K}\ri\}}
{\ds \prod_{j\leq m_1, r \leq m_2} (b_j-a_r)^{k_j \ell_r}}}  
{ \ds \prod_{1\leq j< j'\leq m_1} (b_{j}-b_{j'})^{k_{j}k_{j'}}\prod_{1\leq r<r'\leq m_2} (a_{r}-a_{r'})^{\ell_{r} \ell_{r'}}}  
\label{Fratiocor}
\ee
\ec 
{\bf Proof.} The formula is just a specialization of Thm. \ref{maintheorem} in view of \eqref{standardconjugation}; the only point worth mentioning is that 
 the additional terms in lines \eqref{118} \eqref{119} are absent because on each contour of $\Sigma$ the jump matrices are of the form $\1 + $triangular, and hence the matrix under trace is strictly triangular (upper or lower) so that the integrands vanish identically. \QED
 \begin{example}
 In  the case where either $\b=\emptyset \neq \a$ or $\a = \emptyset \neq \b$  and all multiplicities are one, we obtain, respectively,
\bea
\label{Darbouxdiff-}
\pa \ln F^{(\a,\emptyset)}  &=& \pa \ln  F^{(\emptyset,\emptyset)} + \pa \ln \det G_{
\tiny \le\{\hspace{-7pt}
\begin{array}{c}
\a \ \ \ \ \ \ \ \{\infty \}\cr
  \{(E_{11})^m\}  \{mE_{11}\} 
\end{array}\hspace{-7pt}
\ri\}} - \pa \ln \Delta (\a)\ ,\ \ \\
\label{Darbouxdiff+}
\pa \ln F^{(\emptyset,\b)}  &=& \pa \ln F^{(\emptyset, \emptyset)} + \pa \ln \det G_{
\tiny \le\{\hspace{-7pt}
\begin{array}{c}
\{\infty \}\ \ \ \ \ \ \ \b   \cr
  \{mE_{11}\}  \{(E_{11})^m\} 
\end{array}\hspace{-7pt}
\ri\}} - \pa \ln \Delta (\b)\ ,
\eea
where $\Delta (\a) = \prod_{j<\ell} (a_j-a_\ell)$ (the Vandermonde determinant) and $(E_{11})^m$ in the symbol above simply stands for the sequence of $m$ copies of $E_{11}$. The relevant characteristic matrices are
\be\label{matrixG}
	G_{
\tiny \le\{\hspace{-7pt}
\begin{array}{c}
\a \ \ \ \ \ \ \ \{\infty \}\cr
  \{(E_{11})^m\}  \{mE_{11}\} 
\end{array}\hspace{-7pt}
\ri\}}= \le[\res_{z = \infty} \frac{\left[z^{i - 1}\Gamma^{-1}(a_j)\Gamma(z)\right]_{1,1}}{z - a_j}\ri]_{i,j\leq m} 
	G_{
\tiny \le\{\hspace{-7pt}
\begin{array}{c}
\{\infty \}\ \ \ \ \ \ \ \b   \cr
  \{mE_{11}\}  \{(E_{11})^m\} 
\end{array}\hspace{-7pt}
\ri\}} = \le[\res_{z = \infty} \frac{\left[z^{i - 1}\Gamma^{-1}(z)\Gamma(b_j)\right]_{1,1}}{b_j-z}\ri]_{i,j\leq m}
\ee
The formulas \eqref{Darbouxdiff+}, \eqref{Darbouxdiff-} imply at once that 
\bea\label{ratiotaufunctions}
\frac{F^{(\a,\emptyset)}}{F^{(\emptyset,\emptyset)}}  =C \frac{\det G_{
\tiny \le\{\hspace{-7pt}
\begin{array}{c}
\a \ \ \ \ \ \ \ \{\infty \}\cr
  \{(E_{11})^m\}  \{mE_{11}\} 
\end{array}\hspace{-7pt}
\ri\}}}{\Delta(\a)}, \quad \frac{F^{(\emptyset,\b)}}{F^{(\emptyset,\emptyset)}}   =C  \frac{\det G_{
\tiny \le\{\hspace{-7pt}
\begin{array}{c}
\{\infty \}\ \ \ \ \ \ \ \b   \cr
  \{mE_{11}\}  \{(E_{11})^m\} 
\end{array}\hspace{-7pt}
\ri\}}}{\Delta(\b)}
\eea
for some constant $C$ that does not depend on $\a,\b$ or any of the parameters of the RHP; it may only depend on $m$. Clearly this constant must be fixed by some other consideration and we will see that it turns out to be $C=1$, in the cases we will consider. 
 \end{example}
\begin{remark}
Suppose now, for simplicity, that $\#\mathcal A = \#\mathcal B = m$ and that all the $a_i,b_j$ are distinct.  Denote with $\Xi(\a,\b)$ the ratio
$$
\Xi(\a,\b) = \frac{F^{(\a,\b)}}{F^{(\emptyset,\emptyset)}}.
$$ 
When $m = 1$ we have
$$
\Xi_1(a,b) = -C\left(\Gamma^{-1}(a)\Gamma(b)\right)_{11}
$$
where, here and below, $C$ denotes an absolute constant,i.e., independent of $\a,\b,\vtau,\s$. More generally,
\be
	\Xi_m(\a,\b) = C\det\left[\frac{(\Gamma^{-1}(a_i)\Gamma(b_j))_{11}}{a_i - b_j}\right] \frac{\prod_{i,j = 1}^m(b_i-a_j)}
	{\Delta(\a)\Delta(\b)};
\ee
combining the two we get
\be\label{reproducingKP}
	\Xi_m(\a,\b) = C\det\left[\frac{\Xi_1(a_i,b_j)}{b_j - a_i}\right] \frac{\prod_{i,j = 1}^m(b_i-a_j)}{\Delta(\a)\Delta(\b)}.
\ee
Essentially, this means that the value of $\Xi$ at a single point $(a,b) \in \C^2$ is enough to reconstruct the value of $\Xi$ at an arbitrary point $(\a,\b) \in \C^{2N}$.

Exactly the same reproducing formula holds for an arbitrary KP tau function, when written in Miwa variables (see for instance \cite{KonAl}, \cite{ASvM} and \cite{ZabrodinHirota}; in this last reference in particular it is explained that \eqref{reproducingKP}, when the language of free fermions is used, is nothing but Wick's theorem).\\
In other words, the quantity $\Xi(\a,\b)$, thought as a formal series around infinity with respect to the variables $\a,\b$, satisfies the KP bilinear equations (in Miwa variables)
\be\label{bilinearMiwa}
		\oint \frac{dz}{2\pi i } \prod \frac{ ( 1 -  z/b_k) (1- z/a'_k ) }{ ( 1 -  z/b'_k) (1- z/a_k ) }\, \Xi(\a;\b,z) \, \Xi(\a',z;\b') = 0.
\ee
where here we have $\#(\a) = \#(\b,z) = \#(\a',z) = \#(\b')$. We underline that this connection with KP equations, which we find interesting and rather curious, is  formal; KP equations will be used also below in a more effective way to (re)-deduce a PDE for the gap probability associated to the B-B-P distribution.
\end{remark}
We shall show that the matrices \eqref{matrixG} can be written in a  form which involves derivatives in the endpoint variables.

Suppose that $\Gamma(z,s):= \Gamma^{(\emptyset,\emptyset)}(z,s)$ is known.
We need the following lemma. 
 \begin{lemma}\label{lemma1}
 	Let us denote generically with $s$ any of the endpoint variables $s_\ell^{(i)}$; then the following formulas hold true:
	\be\label{Gammaderivatives}
		\partial_s\Gamma( z ) =  z \left[\Gamma( z ),\sigma\right] + \left[\Gamma_1,\sigma\right]\Gamma( z )\quad\quad 	\partial_s\Gamma^{-1}( z ) =  z \left[\Gamma^{-1}( z ),\sigma\right] - \Gamma^{-1}( z )\left[\Gamma_1,\sigma\right].
	\ee
	where 
\be
\sigma:= \frac{1}{z}\pa_{s}\, T(z) \ ; \ \ \ \Gamma(z) = \1  + \frac {\Gamma_1}{z} + \mathcal O(z^{-2})\ ,\ \ \ z\to\infty.
\ee
 \end{lemma}
{\bf Proof.} We observe that, since the jumps associated to $\Psi( z ) := \Gamma( z ){\rm e}^{T( z )}$ are constant, 
$$U := (\partial_s\Psi( z ))\Psi^{-1}( z )$$
is an entire function (no jumps) bounded by a polynomial of degree one; using the asymptotic expansion of $\Psi( z )$ at infinity we obtain $U( z ) =  z \sigma + [\Gamma_1,\sigma]$. The first formula in \eqref{Gammaderivatives} is obtained from $\partial_s\Psi( z ) = U( z ) \Psi( z )$ simply plugging in the definition of $\Psi( z )$, while the second one is obtained from the first one with standard manipulations. \QED

In order to establish connection with the existing literature \cite{BaikPformulas} we need some more informations on the matrices \eqref{matrixG} and their properties. Consider the following sequence (in $\ell$) of $(N+1)\times (N+1)$ matrices ($(N+1)$ is  the size of the RHP \eqref{genericRH1MT}; the dependence on the other parameters $\vtau, \s$ are understood throughout)
\be
\label{blockmatrixG-}
	\mathcal G_\ell^{-}  = \mathcal G_\ell^{-} {(a)} :=  \res_{ z  = \infty} \frac{ z ^{\ell -1}}{ z  - a}\Gamma^{-1}(a)\Gamma(z), \quad \; \ell = 1,2,\dots, 
\ee 
\be
\label{blockmatrixG+}
	\mathcal G_\ell^{+} = \mathcal G_\ell^{+} {(b)} := \res_{ z  = \infty} \frac{ z ^{\ell -1}}{ b-z}\Gamma^{-1}(z)\Gamma(b), \quad \; \ell = 1,2,\dots.
\ee 
\bp\label{propG}
Let $s$ and $\sigma$ be as in Lemma \ref{lemma1}; then the following formulas hold true:
\bea
\ds	\partial_s\, \mathcal G^{-}_{\ell} = a[\mathcal G^{-}_{\ell},\sigma]  +\mathcal G^{-}_{1} [\Gamma_\ell,\sigma] 
	\qquad &\&  
\ds 	\partial_s\, \mathcal G^{+}_{\ell} = b[\mathcal G^{+}_{\ell},\sigma] - [\sigma,\Gamma_\ell^{-1}]\mathcal G^{+}_{1}
\label{firstformula}
	\\[10pt]
\ds	\mathcal G^{-}_{\ell+1}=  a  \mathcal G^{-}_{\ell} + \mathcal G^{-}_{1} \Gamma_{\ell}
\qquad &\&
\ds	\mathcal G^{+}_{\ell+1} =  b  \mathcal G^{+}_{\ell} + \Gamma^{-1}_\ell \mathcal G^{+}_{1}\label{secondformula} 
\eea
Equivalently, the following recursion relation holds, for the entries $(\mu,\nu)$:
\bea
\label{recursionrelationa}
	\Big(\pa_s + a(\pm 1-h_{\mu,\nu})\Big)(\mathcal G_{\ell}^{-})_{\mu,\nu} = \pm  (\mathcal G_{\ell+1}^{-})_{\mu,\nu} +  \Big( \mp\mathcal G^{-}_{1} \Gamma_\ell +\mathcal G^{-}_{1} [\Gamma_\ell,\sigma]\Big)_{\mu,\nu},\\
\label{recursionrelationb}
	\Big(\pa_s + b(\mp 1-h_{\mu,\nu})\Big)(\mathcal G_{\ell}^{+})_{\mu,\nu} =
	\mp (\mathcal G_{\ell+1}^{+})_{\mu,\nu} 
	+ \Big(\mp \Gamma^{-1}_\ell\mathcal G^{+}_{1}- 
[\sigma,\Gamma^{-1}_\ell]\mathcal G^{+}_{1} \Big)_{\mu,\nu},	
\eea
where the coefficients $\{h_{\mu,\nu}\}$ are defined by $[E_{\mu,\nu},\sigma] = h_{\mu,\nu}E_{\mu,\nu}$ (or also $[\sigma,E_{\nu\mu}] = h_{\mu\nu}E_{\nu\mu}$), and\\ $\Gamma^{\pm 1}_\ell :=- \res_{z=\infty} z^{\ell-1} \Gamma^{\pm 1}(z)\d z$. 
\ep
{\bf Proof.} The formula \eqref{firstformula} for the $(\a)$ case is proven 
by differentiating the definition of $\mathcal G^{-}_{\ell}(a)$ using Leibnitz rule; the derivative then involves Lemma \ref{lemma1} and the computation proceeds by straightforward simplification. Skipping some intermediate steps we have  the following chain of identities:
\bea
\pa_s\, (\mathcal G_{\ell}^{-})_{\mu,\nu} &=&  \res_ { z =\infty}\frac {  z ^{\ell-1}}{ z -a} 
\pa_s\, \le(  \Gamma^{-1}(a)\Gamma( z )\ri)_{\mu,\nu}
\nonumber \\
&=& \res_ { z  = \infty}\frac {  z  ^{\ell-1}}{ z  - a}  \tr \Bigg(a \big[\Gamma^{-1}(a), \sigma\big] \Gamma( z ) E_{\nu\mu} + 
\Gamma^{-1}(a)  z  \big[\Gamma( z ), \sigma\big] E_{\nu\mu} \Bigg) \nonumber \\
&=& \res_ { z  = \infty}\frac {  z ^{\ell-1}}{ z -a}  \tr \le({a}  \Gamma^{-1}(a)\Gamma( z )h_{\mu\nu} E_{\nu,\mu}  +  {( z -a)}  
\Gamma^{-1}(a)[\Gamma( z ), \sigma] E_{\nu,\mu} \ri)\nonumber \\
&=& h_{\mu\nu} {a} (\mathcal G_{\ell}^{-})_{\mu,\nu} - \le (   
\Gamma^{-1}(a)[\Gamma_\ell, \sigma] \ri)_{\mu,\nu}
= \le( a \, [ \mathcal G_{\ell}^{-} ,\sigma] -\Gamma^{-1}(a)[\Gamma_{\ell}, \sigma]  \ri)_{\mu,\nu} \nonumber
\eea
The case of $\pa_s\, (\mathcal G_{\ell}^{+})_{\mu,\nu}$ is handled in a similar fashion.
For \eqref{secondformula} we observe that, from the very definition
\bea
\mathcal G_{\ell+1}^{-} = \res_{z=\infty} \frac {z^{\ell}}{z-a} \Gamma^{-1}(a)\Gamma(z) =  
 \res_{z=\infty} \frac {z^{\ell-1}(z-a+a)}{z-a} \Gamma^{-1}(a)\Gamma(z) =  a\, \mathcal G_{\ell}^{-} - \Gamma^{-1}(a)\Gamma_\ell
 = a\, \mathcal G_{\ell}^{-} +\mathcal G_{1}^{-}\Gamma_\ell\\
 \mathcal G_{\ell+1}^{+} = \res_{z=\infty} \frac {z^{\ell}}{b-z} \Gamma^{-1}(z)\Gamma(b) =  
 \res_{z=\infty} \frac {z^{\ell-1}(z-b+b)}{b-z} \Gamma^{-1}(z)\Gamma(b) =  b\, \mathcal G^{(+)}_{\ell} + \Gamma^{-1}_\ell\Gamma(b)
 = b\, \mathcal G^{(+)}_{\ell} + \Gamma^{-1}_\ell \mathcal G^{(+)}_{1} 
\eea
Finally, equation \eqref{recursionrelationa} comes from \eqref{firstformula} and \eqref{secondformula} by adding or subtracting, observing also  that
$[\mathcal G_{\ell}^{\pm},\sigma]_{\mu,\nu} = h_{\mu,\nu}(\mathcal G_{\ell}^{\pm})_{\mu,\nu}.$ \QED

\bc\label{maincor}
Given $F^{(\a,\b)}(\vtau,\s)$ associated to the Riemann--Hilbert problem \ref{genericRH1MT}  for $\a,\b = \emptyset$ respectively, we have 
\be\label{detGb}
	\partial_{s_1^{(1)}} \ln \frac{F^{(\b,\emptyset)}}{F^{(\emptyset,\emptyset)}} = \pa_{s_1^{(1)}} \ln \frac 1 {\Delta (\b)} \det \Big[(-\pa_{s_1^{(1)}} + b_j)^{\ell - 1} f(b_j) \Big]_{\ell,j = 1}^m
\ee
\be\label{detGa}
	\partial_{s_1^{(1)}} \ln \frac{F^{(\emptyset,\a)}}{F^{(\emptyset,\emptyset)}} = \pa_{s_1^{(1)}} \ln \frac 1 {\Delta (\a)} \det \Big[(\pa_{s_1^{(1)}} + a_j)^{\ell - 1} g(a_j) \Big]_{\ell,j = 1}^m
\ee
with $f(b_j) := \Gamma_{1,1}(b_j),\quad g(a_j) := \le(\Gamma\ri)^{-1}_{1,1}(a_j)$.
\ec
{\bf Proof.}
We take into consideration the case of $\tau^{(\a)}$.  In this case the characteristic matrix is $[(\mathcal G_{\ell}^{-}(a_j))_{11}]_{j, \ell \leq m} $. We thus have to prove that
\bea\label{detG1}
	\det[(\mathcal G_{\ell}^{-}(a_j))_{11}]_{j, \ell \leq m}  = C \det \Big[(\pa_{s_1^{(1)}} + a_j)^{\ell-1} g(a_j)\Big]_{\ell,j = 1}^{m}.
\eea
where $C$ is a constant independent of the deformation parameters. In order to prove \eqref{detG1} we observe that, in this particular case, $\sigma = \frac{1}{N+1}(N,-1,\ldots,-1)$ and the equation \eqref{recursionrelationa} (with the plus sign) reads as
$$
(\pa_{s_1^{(1)}} + a_j)(\mathcal G_{\ell}^{-}(a_j))_{11} = (\mathcal G_{\ell+1}^{-}(a_j))_{11} -(\mathcal G_{\ell}^{-}(a_j))_{11}(\Gamma_\ell)_{11}.
$$ 
On the other hand $\mathcal G_{1}^{-}(a_j) = -(\Gamma^{-1}(a_j))_{1,1} = -g(a_j)$, hence the statement is proven changing the sign of the first column of $G$ and then performing linear transformations between the columns. The second case is handled similarly. \QED

The goal of the following sections is to show how  specializations of the setup in the previous Sec. \ref{sectGP} cover several instances already appeared in the literature and then also allows us to obtain interesting generalizations.
\subsection{The Baik-Ben Arous-Pech\'e distribution and Baik's formula}
\label{examples}
\begin{wrapfigure}{r}{0.3\textwidth}
\resizebox{0.30\textwidth}{!}{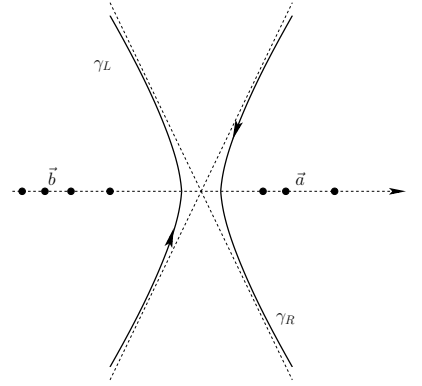}
\caption{\footnotesize{Contours $\gamma_{R/L}$ and position of the points $\a,\b$.}}
\label{contours}
\end{wrapfigure}

Our first example is the  B-B-P distribution \cite{BaikBenArousPeche}, i.e., in our notation
\be\label{BBPdistribution}
	F_{BBP}^{(\emptyset,\b)}(s) := \det\left(\Id - K_\Ai^{(\b)}\chi_{[s,\infty)}\right),
\ee
where 
\be\label{BBPkernel}
	K^{(\emptyset,\b)}_{\Ai}(x,y) := \frac{1}{(2\pi i)^2} \int_{\gamma_L}\!\!\!dw\int_{\gamma_R}\!\!\!dz \frac{{\rm e}^{\frac{z^3}3 - \frac{w^3}3 - zx + wy}}{w-z}\frac {B(z)}{B(w)},
\ee
$B(z) := \prod_{k = 1}^m\left({z - b_k}\right)$ 
and $\gamma_{L/R}$ are depicted in Figure \ref{contours}, together with their reciprocal position with respect to the parameters $\b := \{b_1,\ldots, b_m\}$. The relevant Riemann--Hilbert problem, in this case, is a $(2\times 2)$ problem reading as follows:

\begin{problem}\label{RHAiry}
Find the sectionally analytic function $\Gamma^{(\emptyset,\b)}( z ) \in \mathrm{GL}(2,\C)$ on $\C/\left\{\gamma_R \cup \gamma_L\right\}$ such that
\be\label{RHAiryeq}
	\left\{\begin{array}{ll}
	\Gamma_+^{(\emptyset,\b)}( z ) = \Gamma_-^{(\emptyset,\b)}( z )\left[\begin{array}{cc}
								1 & -{\rm e}^{\frac{ z ^3}3 - s z }B(z) \chi_R\\
								\\
								 -{\rm e}^{-\frac{ z ^3}3 + s z }B(z)^{-1}\chi_L & 1
						\end{array}\right]&\\
	&\\
	\Gamma^{(\emptyset,\b)}( z ) \sim \Id + \mathcal O( z ^{-1}), \quad  z  \rightarrow \infty.&
	\end{array}\right.					 
\ee
\end{problem}
Observe that, for $m = 0$, this is precisely the Riemann--Hilbert problem associated to the Hasting--Mc Leod solution of the Painlev\'e II equation. As corollary of the general theory exposed in the previous sections, we obtain that, up to $s$--independent terms, $F_{BBP}(s)$ coincide with the isomonodromic tau function associated to \ref{RHAiry}. Moreover we also obtain the following formula, already obtained by Baik in \cite{BaikPformulas} (with slightly different notations).

\bc[\cite{BaikPformulas}]
\label{corBaik}
Let $f( z ) := \Gamma^{(\emptyset,\emptyset)}_{1,1}( z )$ be the $(1,1)$--entry of the solution to \ref{RHAiry} for $\b = \emptyset$. Then, for an arbitrary set of parameters $\b$, we have
\be\label{BaikAiryformula}
	F_{BBP}^{(\b)}(s) = F_{TW}(s) \frac{\det \Big((-\pa_s + b_j)^{\ell - 1} f(b_j)\Big)_{\ell,j = 1}^m}{\Delta(\b)},
\ee
where $F_{TW}(s)$ is the standard Tracy--Widom distribution.
\ec
\emph{Proof:} From \eqref{detGb} and Cor. \ref{maincor} we have, integrating once, that 
$$F_{BBP}^{(\b)}(s) = C F_{TW}(s) \frac{\det \Big((-\pa_s + b_j)^{\ell - 1} f(b_j)\Big)_{\ell,j = 1}^m}{\Delta(\b)}$$ 
and we only need to show that the constant is $C=1$.
 On one hand, both $F_{BBP}^{(\b)}(s)$ and $F_{TW}(s) $ have limit equal to $1$ when $s$ goes to infinity (because of their probabilistic interpretation). Now recall that  $f(z) = \Gamma_{11}(z,s)$,  and it is well known that $\Gamma(z,s) \to \1$ (uniformly) as $s\to +\infty$. Thus $f(z,s)\to 1$ and the determinant of the characteristic matrix tends precisely to the Vandermonde determinant, forcing $C=1$. \QED

\subsection{The Airy kernel with two deformation parameters: generalized Baik's formula}\label{twoparAiry}

In \cite{BorPe,AFvM} the following generalization of \eqref{BBPkernel} (together with its multi--time version) is considered:
\be\begin{array}{ll}
	K^{(\a,\b)}_{\Ai}(x,y):= \nonumber \\\nonumber\\ \nonumber \ds\frac{1}{(2\pi i)^2} \int_{\gamma_L}\!\!\!dw\int_{\gamma_R}\!\!\!dz \frac{{\rm e}^{\frac{z^3}3 - \frac{w^3}3 - zx + wy}}{w-z}
	\frac {B(z)}{B(w)} \frac {A(w)}{A(z)}.
\end{array}\ee
where $B(z) =\prod_{j=1}^m(z-b_j)$ as before and $A(z):= \prod_{k = 1}^q\left({z-a_k}\right)$.
The relevant Riemann--Hilbert problem whose tau function coincides with
\be
	F^{(\a,\b)}(s) := \det(\Id - K^{(\a,\b)}_{\Ai}\chi_{[s,\infty)})
	\label{F343}
\ee
is the following:
\begin{problem}\label{RHAiry2}
Find the sectionally analytic function $\Gamma^{(\a,\b)}( z ) \in \mathrm{GL}(2,\C)$ on $\C/\left\{\gamma_R \cup \gamma_L\right\}$ such that
\be
\label{RHAiryeq2}
	\left\{\begin{array}{ll}
	\Gamma_+^{(\a,\b)}( z ) = \Gamma_-^{(\a,\b)}( z )\left[\begin{array}{cc}
								1 & -{\rm e}^{\frac{ z ^3}3 - s z }\ds
								\frac {B(z)}{A(z)} \chi_R\\
								\\
								 -{\rm e}^{-\frac{ z ^3}3 + s z }\ds\frac {A(z)}{B(z)}\chi_L & 1
						\end{array}\right]&\\
	&\\
	\Gamma^{(\a,\b)}( z ) \sim \Id + \mathcal O( z ^{-1}), \quad  z  \rightarrow \infty.&
	\end{array}\right.					 
\ee
where $A(z) := \prod_{k = 1}^q(z - a_k),\ 
B(z) := \prod_{k = 1}^m (z-b_k)$.
\end{problem}
	Given $\Gamma$ the solution of the Riemann-Hilbert problem \ref{RHAiry2} for $(\a) = (\b) = \emptyset$ ($\Gamma$ is, in other word, the solution associated to the Hastings-McLeod solution of the Painlev\'e II equations), there are two distinct way to obtain $\Gamma^{(\a,\b)}$ through Darboux transformations.\\
The first one consists in conjugating the jump for the ``bare problem'' with  the diagonal matrix $D := {\rm diag} (A(z), B(z))$ (hence we slightly deviate from what described in formula \eqref{standardconjugation}). In this case, the relevant characteristic matrix will be $G := G_{
{\le\{\tiny
\begin{array}{cc}
\{(\a) ,(\b)\} & \{\infty \} \\
\{(E_{11})^q(E_{22})^m\} & \{(qE_{11}+mE_{22})\}
\end{array}
\ri\}}}$ written in block form as
\be
G := \res_{ z  = \infty} \left[\begin{array}{c|c}
		&\\
		 \ds\frac{\Big[ z ^\ell\Gamma^{-1}(a_j)\Gamma( z )\Big]_{1,1}}{ z -a_j} &  \ds\frac{\Big[ z ^\ell\Gamma^{-1}(a_j)\Gamma( z )\Big]_{1,2}}{ z -a_j} \\
		 \\
		 \hline
		 \\
	\ds\frac{\Big[ z ^\ell \Gamma^{-1}(b_j)\Gamma( z )\Big]_{2,1}}{ z -b_j}	&  \ds\frac{\Big[ z ^\ell\Gamma^{-1}(b_j)\Gamma( z )\Big]_{2,2}}{ z -b_j}\\
	&
	\end{array}\right] = \le[
	\begin{array}{c|c}
	\mathcal G_\ell^{-}(a_j)_{1,1} &\mathcal G_\ell^{-}(a_j)_{1,2} \\
	\\
	\hline\\
	\mathcal G_\ell^{-}(b_j)_{2,1} &\mathcal G_\ell^{-}(b_j)_{2,2}  
	\end{array}
	\ri]	
	\label{344}
\ee
In each block of the formula above, the index $j$ is the row-index, the index $\ell$ the column
and  the upper-left block is of size $(q\times q)$ and the lower-right $(m\times m)$. Then 
\bp
	Denote with $g_{ij}(z)=(\Gamma^{-1})_{ij}(z)$, where $\Gamma$ is the  solution of the Riemann-Hilbert problem for the Hasting Mc--Leod solution of Painlev\'e II (i.e. \ref{RHAiry2} for $(\a) = (\b) = \emptyset$). Then, for arbitrary sets of parameters $\a$ and $\b$, 
\be\label{blockBaik}
F^{(\a,\b)}(s) = F_{TW}(s)\frac{\det\left[\begin{array}{c|c}
&\\
 {\ds \big(\pa_s + a_j\big)^{\ell -1} g_{11}(a_j)} \atop { 1\leq \ell,j \leq q} &   
 {\ds 
 {(-\pa_s)}^{\ell -1} g_{12}(a_j)}\atop {1\leq \ell \leq m; j\leq q }\\
 \\
 \hline
 \\
{\ds \pa_s^{\ell -1} g_{21}(b_j)}\atop{1 \leq\ell \leq q; j\leq m}	& 
{\ds \big(
{-\pa_s} + b_j\big)^{\ell -1} g_{22}(b_j)}\atop {1 \leq \ell,j\leq m}\\
&
\end{array}\right]}{\Delta(\a)\Delta(\b)}.
\ee
\ep
{\bf Proof.}
According to Corollary \ref{corFratio} specialized to this case, the ratio $ F^{(\a,\b)}(s) / F_{TW}(s)$ is 
\be       
\frac{F^{(\a,\b)}}{ F_{TW}}   = C \frac {\det G } 
{ \ds\Delta(\a) \Delta(\b)} 
\ee
with a constant that cannot depend on any of $s, \vec a, \vec b$. We first manipulate $\det G$ into the determinant that appears in \eqref{blockBaik}, and then show that $C=1$. 

First of all, the matrix $\sigma$ appearing in Prop. \ref{propG} is $\frac 12 \sigma_3  = {\rm diag} (\frac 1 2, -\frac 1 2)$. 
Next we look at the entries $(1,1)$ and $(2,1)$ of \eqref{recursionrelationa} with the $+$ sign,  and obtain the following recursion relations for $\mathcal G^{-}_\ell$
\bea
(\pa_{s} + a_j)(\mathcal G_{\ell}^{-})_{11}(a_j) =
  (\mathcal G_{\ell + 1}^{-})_{11}(a_j) - (\mathcal G_{1}^{-})_{11}(a_j)(\Gamma_\ell)_{11}\\
\pa_{s} (\mathcal G_{\ell}^{-})_{21}(b_j) = 
(\mathcal G_{\ell + 1}^{-})_{21}(b_j) - (\mathcal G_{1}^{-})_{21}(b_j)(\Gamma_\ell)_{11}
\eea
Looking now at the entries $(1,2), (2,2)$ of  \eqref{recursionrelationa} with the $-$ sign we obtain 
\bea
(\pa_{s} - b_j)(\mathcal G_{\ell}^{-})_{22}(b_j) = 
 -(\mathcal G_{\ell + 1}^{-})_{22}(b_j) + (\mathcal G_{1}^{-})_{22}(b_j)(\Gamma_\ell)_{22}\\
\pa_{s} (\mathcal G_{\ell}^{-})_{12}(a_j) = 
-(\mathcal G_{\ell + 1}^{-})_{12}(a_j) + (\mathcal G_{1}^{-})_{12}(a_j)(\Gamma_\ell)_{22}
\eea
Rearranging and simplifying we find 
\bea
\begin{array}{c|c}
\ds  (\mathcal G_{\ell + 1}^{-})_{11}(a_j)  = (\pa_{s} + a_j)(\mathcal G_{\ell}^{-})_{11}(a_j)+ (\mathcal G_{1}^{-})_{11}(a_j)(\Gamma_\ell)_{11} & \ds (\mathcal G_{\ell + 1}^{-})_{12}(a_j)  = -\pa_{s} (\mathcal G_{\ell}^{-})_{12}(a_j) 
+ (\mathcal G_{1}^{-})_{12}(a_j)(\Gamma_\ell)_{22}
\\
\ds (\mathcal G_{\ell + 1}^{-})_{21}(b_j)   =\pa_{s} (\mathcal G_{\ell}^{-})_{21}(b_j) +  (\mathcal G_{1}^{-})_{21}(b_j)(\Gamma_\ell)_{11}
& 
\ds (\mathcal G_{\ell + 1}^{-})_{22}(b_j)  =-(\pa_{s} - b_j)(\mathcal G_{\ell}^{-})_{22}(b_j) + (\mathcal G_{1}^{-})_{22}(b_j)(\Gamma_\ell)_{22}
\end{array}
\label{rec3}
\eea
Using \eqref{rec3} repeatedly on the columns of \eqref{344} we see that we obtain a triangular transformation that leaves the determinant unchanged and leads to the determinant in the numerator of \eqref{blockBaik}.

We still need to show that $C=1$; we use the fact that $\Gamma(z;s)\to \1$ (uniformly on $\C \mathbb P^1$) as $s\to+\infty$ so that $g_{ij}(a;s)\to \delta_{ij}$. Then, taking the limit of both sides of \eqref{blockBaik} and comparing we find the result. 
\QED

The formula \eqref{blockBaik} is ostensibly a very expectable generalization of the formula \eqref{BaikAiryformula} originally found by Baik. 
A second way would be to alternatively use the standard conjugation by the $D$ written in \eqref{standardconjugation}, as we will do for the multi--time case below. This takes the simplest form in the case $q=m$ (same number of outliers and inliers); 
in this case the jumps of the RHP \ref{RHAiry2} are seen to be obtained from the ``bare'' jumps by conjugating with $D(z)  = {\rm diag} \le(
\frac {A(z)}{B(z)}, 1
\ri)$ and thus leads Thm .\ref{maintheorem} yields immediately
\be
F^{(\a,\b)}(s) =C  F_{TW}(s)\frac {\ds \det\left[\frac{(\Gamma^{-1}(a_i)\Gamma(b_j))_{11}}{a_i - b_j}\right]\prod_{i,j = 1}^m(a_i-b_j)}{\ds  \Delta(\a)\Delta(\b)}
\label{232}
\ee
with $C$ --again-- an absolute constant. Taking the limit $s\to +\infty$ the reader notices that $(\Gamma^{-1}(a_j;s) \Gamma(b_j;s))_{11}\to 1$, at which point the determinant in \eqref{232} becomes the Cauchy determinant that cancels exactly the other terms. Thus $C=1$ as before.

\subsubsection{A nonlinear PDE}
We now show how to deduce a nonlinear PDE for the Fredholm determinant $F^{(\a,\b)}(s)$ \eqref{F343} via the RHP \ref{RHAiry2}. 

To this end we need to allow the phase function $\theta(z)=  z^3/3-sz$ to depend on an infinite sequence of ``times'' $t_j$. 
Thus, let $ \Gamma^{( \a,\b)} (z;\un t)$ be the solution of the RHP \ref{RHAiry2} but with the substitution of the phase 
\be
\theta(s;z)=\frac {z^3}3 - s z \mapsto \sum_{j=1}^\infty t_j z^j = \theta(\un t;z)\ ,
\ee
By the substitution we mean that ${\rm e}^{\frac {z^3}3 - s z}$ is replaced by a function $g(z;\un t) = {\rm e}^{\sum_{j=1}^\infty t_j z^j}$
that is invertible on the contours and decays at infinity along said contours. This step may be regarded as formal but in the end of the computation we will set $t_j=0$ for all $j\geq 4$. With this preparation we can formulate 
\bt
\label{nonlinPDE}
Let $\mathcal F(\a,\b,s) = \ln F^{(\a,\b)}(s)$ (see \eqref{F343}) and denote
\be 
\mathbb T:= \sum \pa_{a_j} +\sum \pa_{b_j}\ ,\qquad \mathbb E:= \sum a_j\pa_{a_j} +\sum b_j\pa_{b_j}
\ee
Then the following nonlinear PDE  is satisfied:
\bea
{\frac {\pa^{4} \mathcal F}{\pa v^{4}}}
+
2\,{\frac {\pa \mathcal F}{\pa v}}
-
4 v  {\frac {\pa^{2} \mathcal F}{\pa { v}^{2}}} 
+
6\, \left( {\frac {\pa^{2} \mathcal F}{\pa {v}^{2}}} \right) ^{2}
+
3 \mathbb T^2 \mathcal F
+
4\mathbb E \frac{\pa \mathcal F}{\pa v} = 0
\label{nnPDE}
\eea
\et
{\bf Proof.}
We shall show that the equation \eqref{nnPDE} is nothing but the ordinary KP equation in disguise.
We start by considering the matrix-valued function
\be
H(z;\un t, \un {\wt t} , \a,\b):= \Gamma^{( \a,\b)} (z;\un t) {\rm e}^{\sum_{j} (t_j-\wt t_j)z^jE_{11}} (\Gamma^{( \a,\b)} (z;\un {\wt t}))^{-1} 
\ee
We let the reader the simple verification that $H$ 
is an {\em analytic} (actually, entire)  function of $z$; by the Cauchy residue theorem thus we have the identity 
\be
0_{2\times 2} \equiv \oint_{|z|=R} H(z;\un t, \un {\wt t} , \a,\b)\d z.
\ee
Taking the $(1,1)$ entry of the above yields 
\be
0\equiv  \oint_{|z|=R} \Gamma^{( \a,\b)}_{11}  (z;\un t) {\rm e}^{\sum_{j} (t_j-\wt t_j)z^j} (\Gamma^{( \a,\b)} (z;\un {\wt t}))^{-1}_{11}   \d z +
 \oint_{|z|=R} \Gamma^{( \a,\b)}_{12}  (z;\un t)  (\Gamma^{( \a,\b)} (z;\un {\wt t}))^{-1}_{21}   \d z 
\ee
The integrand of the second integral behaves as $z^{-2}$ uniformly in each sector and hence, in the limit $R\to \infty$ it vanishes. Thus
\be
0\equiv  \lim_{R\to\infty} \oint_{|z|=R} \Gamma^{( \a,\b)}_{11}  (z;\un t) {\rm e}^{\sum_{j} (t_j-\wt t_j)z^j} (\Gamma^{( \a,\b)} (z;\un {\wt t}))^{-1}_{11}   \d z
\label{247}
\ee
We now consider the tau function (Fredholm determinant) $F^{(\a,\b)}(s)$  as a function of these additional parameters $\un t$. We next claim that
\bea
F^{(\a,\b)}(\un t + [\z^{-1}]) = F^{(\a,\b)}(\un t) \Gamma^{(\a,\b)}_{11}(\z)
\qquad
F^{(\a,\b)}(\un t - [\z^{-1}]) = F^{(\a,\b)}(\un t) (\Gamma^{(\a,\b)})^{-1}_{11}(\z)\\{}
[\z^{-1}] := \le(\frac 1 \z , \frac 1 {2 \z^2}, \frac 1{3\z^3}, \dots,\ri).\label{362}
\eea
which is the so--called "Miwa shift" in the lore of integrable systems.
Eq. \eqref{362}  is precisely an instance of Thm. \ref{maintheorem} because the Miwa shift has the only meaning of the replacement $A(z)\mapsto  (z-\z)A(z)$ in the RHP \ref{RHAiry2}: this is due to the trivial but far reaching oservation that 
\be
\exp \theta( \un t \pm [\zeta^{-1}];z) = \exp \le(\theta(\un t;z) \pm \sum_{j=1}^\infty \frac {z^j}{j\z^j}\ri) =  (1 - \frac z \z)^{\mp 1}\exp \le(\theta(\un t;z) \ri)
\ee
 Thus the identity \eqref{247} becomes 
\be
0\equiv \res_{\z=\infty} \oint_{|z|=R} F^{( \a,\b)}  (\un t + [\z^{-1}] ) {\rm e}^{\sum_{j} (t_j-\wt t_j)\z^j} F^{( \a,\b)} (\un {\wt t} - [\z^{-1}])   \d \z
\ee
where the residue is the formal residue (really, the limit \eqref{247}). This is an instance of Hirota bilinear equations for a KP tau function. This simply means that the function $F^{(\a,\b)}(\un t)$ satisfies an infinity of equations in the variables $\un t$; these equations involve an increasing number of derivatives with respect to the parameters $t_j$ with  higher and higher index. Only the very first one (which is, incidentally, a bilinearized version of the historical original KP equation) involves only $t_1,t_2,t_3$ and reads 
\bea
&\&(D_1^4 - 4D_1D_3 + 3D_2^2)F^{(\a,\b)}\cdot F^{(\a,\b)} =0\label{KP}\\
&& D_\mu^k F^{(\a,\b)}\cdot F^{(\a,\b)}:= \frac {\d^k}{\d h^k} F^{(\a,\b)}(\dots,t_\mu + h,\dots) F^{(\a,\b)}(\dots, t_\mu-h,\dots ).
\eea
We now set $t_4=t_5=\dots =0$.
On the other hand we have the following invariance, a consequence of the group of dilations/translations acting on the RHP
\be
F^{(\a,\b)} (t_1,t_2,t_3) = F^{(\vec \alpha, \vec \beta)}(v,0,1/3) \ ,\qquad v:= \frac{t_2^2-3t_1 t_3 }{3t_3^{\frac 4 3}} ,\ \ \
\alpha_j :=  \frac{3 a_j t_3 + t_2}{3t_3^{\frac 2 3}}\ \ \beta_j :=  \frac{3 b_j t_3 + t_2}{3t_3^{\frac 2 3}}
\ee
This is simply the effect of the transformation 
\be
z = \frac{\z}{(3t_3)^{\frac 1 3} } - \frac {t_2}{3t_3} \ ,\ \ \sum^3 t_j z^j=  \frac {\z^3}3 - v \z + C
\ee
Then the KP equation \eqref{KP} in terms of $\mathcal F:= \mathcal F^{(\a,\b)}:= \ln \F^{(\a,\b)}$ becomes exactly (after trivial algebra) \eqref{nnPDE}.\QED

\subsection{The Airy kernel with outliers}
Now consider the two--times version of the previous example; i.e. let us study
\be\label{BBPMT}
	F^{(\a,\b)}_{\Ai}(s_1,\tau_1,s_2,\tau_2) := \det(\Id - {\bf K}_{\Ai}^{(\a,\b)})\chi_{\vec I}
\ee
where now $\vec I = \Big ( [s_1,\infty) , [s_2,\infty) \Big)$ and ${\bf K}_{\Ai}^{(\a,\b)}$ is the $2 \times 2$ matrix kernel associated to the Airy process with outliers \cite{ADvM,BorPe,AFvM}, i.e. ${\bf K}^{(\a,\b)} = \left(K^{(\a,\b)}_{ij}\right)_{i,j = 1}^2 := \wt K^{(\a,\b)}_{ij} - B\delta_{1i}\delta_{2j},$ with entries given by 
\bea
	\wt K^{(\a,\b)}_{ij}(x,y)&:=&
	\frac{1}{(2\pi i)^2} \int_{\gamma_{R} + \tau_i}\hspace{-10pt} \d w \int_{\gamma_{L} + \tau_j } \d z \frac {{\rm e}^{\frac{z^3}3 - \frac{w^3}3 -zx + wy}}{w + \tau_j - z - \tau_i}\prod_{k = 1}^m\frac{(z- b_k + \tau_i)(w-a_k+\tau_j)}{(w-b_k+\tau_j)(z-a_k+\tau_i)}\label{AKernMT}\\
	B(x,y)&:=& 
	\frac{1}{\sqrt{4\pi(\tau_2-\tau_1)}}{\rm e}^{\frac{(\tau_2-\tau_1)^3}{12}-\frac{(x-y)^2}{4(\tau_2 - \tau_1)} -\frac{(\tau_2 - \tau_1)(x+y)}{2}}.\label{gaussian}
\eea
In order to write  simpler formulas we shall consider only  $\#\mathcal A' = \#\mathcal B' = m$ and with all the $a's$ and $b's$ distinct, but the reader should realize that the scope of the  theory (Thm. \ref{maintheorem} in particular) is completely general.
Moreover  the formula for the case of arbitrary multiplicities can be recovered from the one below by taking limits and using de l'Hopital rule. Shifting the variables of integration in \eqref{AKernMT} we can reduce\footnote{More precisely, the change of integration variables brings \eqref{AKernMT} to the form \eqref{generickernelMT} up to multiplication by ${\rm e}^{x\tau_i - y\tau_j - (\tau_i^3-\tau_j^3)/3}$. This latter, nevertheless, is just a conjugation and it does not affect the Fredholm determinant \eqref{BBPMT}.} this kernel to a kernel of the form \eqref{generickernelMT} with
\be\label{AiryphaseMT}
	\theta_\tau(x;z) := \frac{z^3}{3} - \tau z^2 + (\tau^2 - x)z.
\ee 
Hence \eqref{BBPMT} coincides (up to terms independent by $\s$) with the isomonodromic tau function of the following Riemann--Hilbert problem
\begin{problem}\label{RHAiMT}
Find the sectionally analytic function $\Gamma^{(\a,\b)}( z ) \in \mathrm{GL}(3,\C)$ on $\C/\left\{\gamma_R \sqcup \gamma_L \right\}$ such that
\be\label{RHAiMTeq}
	\left\{\begin{array}{lll}
	\Gamma_+^{(\a,\b)}( z ) = \Gamma_-^{(\a,\b)}( z )M^{(\a,\b)}&\\
	\nonumber&\\
	\Gamma^{(\a,\b)}( z ) \sim \Id + \mathcal O( z ^{-1}), \quad  z  \rightarrow \infty&\\
	\nonumber&\\
	M^{(\a,\b)} := 
	\left[\begin{array}{ccc}
		1 & -{\rm e}^{\theta_{\tau_1}(s; z )}C(z)\chi_R & -{\rm e}^{\theta_{\tau_2}(s_2; z )}C(z)\chi_R\\
		\\
		-{\rm e}^{-\theta_{\tau_1}(s_1; z )}C^{-1}(z)\chi_L & 1 & 0 \\
		\\
		-{\rm e}^{-\theta_{\tau_2}(s_2; z )}C^{-1}(z)\chi_L & -{\rm e}^{-\theta_{\tau_2}(s_2; z )+\theta_{\tau_1}(s_1; z )}\chi_{L} & 1
		\end{array}\right].
	\end{array}\right.					 
\ee
where $C(z) := \ds\prod\frac{(z- b_k)(w-a_k)}{(w-b_k)(z-a_k)}$
\end{problem}

In this case the relevant characteristic matrix associated to the conjugation \eqref{standardconjugation} will be
$G := G_{
{\le\{\tiny
\begin{array}{cc}
  \{ \a \} & \{\b \}\\
\{(E_{11})^m\}& \{(E_{11})^m\}
\end{array}
\ri\}}}$ with entries given by $G_{ij} = \ds \frac{\Big(\Gamma^{-1}(a_i)\Gamma(b_j)\Big)_{11}}{a_i - b_j}$, so that using the general theory we obtain the following:
\bc
Let $\Gamma$ be the solution to \ref{RHAiMT} for $\a = \b = \emptyset$. Then, for an arbitrary set of parameters $\a,\b$ we have
\be\label{BaikAiryformulaMT}
	\frac{F^{(\a,\b)}_\Ai(s_1,\tau_1,s_2,\tau_2)}{F_{\Ai}(s_1,\tau_1,s_2,\tau_2)} = \frac{\det\left[\ds\frac{\left(\Gamma^{-1}(a_i)\Gamma(b_j)\right)_{11}}{a_i - b_j}\right]_{i,j = 1}^m}{\Delta(\a)\Delta(\b)}\prod_{i,j = 1}^m(b_i-a_j)
\ee
\ec
{\bf Proof.} 
Using \eqref{Darbouxdiff-}, one just have to prove that the constant of integration is 0. In order to prove that, one has to send both $s_1,s_2$ to infinity and fact that --again-- $\Gamma(z;s_1,s_2)\to \1_{3}$ can be proven by a simple argument of small-norms operator theory.\QED

\subsection{The Pearcey process with inliers}

For the case of Pearcey, let us start with the $1$--time case considering 
\be
	F^{(\a)}_{\mathrm{P}}(s_1,s_2) := \det(\Id - K^{(\a)}_{\mathrm{P}}\chi_{[s_1,s_2]}),
\ee 
where $K^{(\a)}_{\mathrm P}$ is the integral operator with kernel
\be\label{kernelPearcey}
	K^{(\a)}_{\mathrm P}(x,y) := \frac{1}{(2\pi i)^2} \int_{i\R}\!\!\!dw\int_{\gamma}\!\!\!dz \frac{{\rm e}^{\theta_\tau(x;z)-\theta_\tau(y;w)}}{w-z}\prod_{k = 1}^m\left(\frac{w
	{-}
	a_k}{z
	{-}
	a_k}\right).
\ee
Here $\a := \{a_1,\ldots, a_m\}$ are real parameters and $\theta_{\tau}(x;z) := \frac{z^4}4-\tau\frac{z^2}2-xz$. This kernel (as well as its multi--time version) appeared \footnote{Actually, in \cite{ADvMVA}, this kernel appears with phase $\wt\theta_{\tau}(x;z) = \frac{z^4}4-\tau\frac{z^2}2+xz$, see Remark \ref{wrongsign}.} in \cite{ADvMVA} as the critical kernel describing the so--called Pearcey process with inliers (see the reference for a description of the contours and position of parameters $\a$). The related Riemann--Hilbert problem reads as follows (we denoted $
A
( z ) := \prod_{k = 1}^m ( z  
{-}
 a_k)$):
\begin{problem}\label{RHP}
Find the sectionally analytic function $\Gamma^{(\a)}( z ) \in \mathrm{GL}(3,\C)$ on $\C/\left\{\gamma \sqcup i\R \right\}$ such that
\be\label{RHPearcey}
	\left\{\begin{array}{ll}
	\Gamma_+^{(\a)}( z ) = \Gamma_-^{(\a)}( z )\left[\begin{array}{ccc}
	1 & -{\rm e}^{\theta_\tau(s_1; z )}
	{A}
	^{-1}( z )\chi_\gamma & {\rm e}^{\theta_\tau(s_2; z )}
	{A}
	^{-1}( z )\chi_\gamma
	\\
	 -{\rm e}^{-\theta_\tau(s_1; z )}
	 {A}
	 ( z )\chi_{i\R} & 1 & 0 
	  \\
	 -{\rm e}^{-\theta_\tau(s_2; z )}
	 {A}( z )\chi_{i\R} & 0 & 1
	\end{array}\right]&\\
	\nonumber&\\
	\Gamma^{(\a)}( z ) \sim \Id + \mathcal O( z ^{-1}), \quad  z  \rightarrow \infty.&
	\end{array}\right.					 
\ee
\end{problem}
Using the Corollary \ref {maincor}, and in particular formula \eqref{detGa}, we immediately get
\bc
Let $g( z ) := \Big((\Gamma^{\emptyset})^{-1}\Big)_{1,1}( z )$ be the $(1,1)$--entry of the inverse of the solution to \ref{RHPearcey} for $\a = \emptyset$. Then, for an arbitrary set of parameters $\a$,
\be\label{BaikPearceyformula}
	F^{(\a)}_{\mathrm P}(s_1,s_2) = F_{\mathrm P}(s_1,s_2) \frac{\det \Big((\pa_s 
	{+} 
	a_j)^{\ell - 1} g(
	a_j)\Big)_{\ell,j = 1}^m}{\Delta(
	\a)},
\ee
where $F_{\mathrm P}(s_1,s_2)$ is the gap probability for the Pearcey process on the interval $[s_1,s_2]$.
\ec
\emph{Proof:} Using the Corollary \ref{maincor} and integrating once we obtain that
			$$F_{\mathrm P}^{(\a)}(s_1,s_2) =  C F_{\mathrm P}(s_1,s_2) \det \Big((\pa_s 
			{+}
			 a_j)^{\ell - 1} g(
			 a_j)\Big)_{\ell,j = 1}^m\Delta^{-1}(\a)$$
where $C$ is constant with respect to $\s,\a$. On the other hand, both $F_{\mathrm P}^{(\a)}(s_1,s_2)$ and $F_{\mathrm P}(s_1,s_2)$ are equal to $1$ for $s_1 = s_2$. Using the explicit form of the solution of \ref{RHP} for $s_1 = s_2$ (see formula (4.22) in \cite{BertolaCafasso1}) we prove that $C = 1$. \QED

For completeness we also work out the simplest multi--time case, i.e. the one with two times $\tau_1,\tau_2$ and two intervals $[s_1,s_2], [t_1,t_2]$. The kernel reads
${\bf K}^{(\a)}_{\mathrm P} = \left(K^{(\a)}_{ij}\right)_{i,j = 1}^2 := \wt K^{(\a)}_{ij} - B\delta_{1i}\delta_{2j}$ with entries given by 
\bea
	\wt K^{(\a)}_{ij}(x,y)&:=&
	\frac{1}{(2\pi i)^2} \int_{i\R}\!\!\!dw\int_{\gamma}\!\!\!dz \frac{{\rm e}^{\theta_{\tau_i}(x;z)-\theta_{\tau_j}(y;w)}}{w-z}\prod_{k = 1}^m\left(\frac{w
	{-}
	a_k}{z
	{-}
	a_k}\right) \label{PKern}\\
	B(x,y)&:=& 
	\frac{1}{\sqrt{2\pi(\tau_2-\tau_1)}}{\rm e}^{-\frac{(x-y)^2}{2(\tau_2 - \tau_1)}}\label{gaussianP}
\eea
and we consider its Fredholm determinant restricted to the collection of intervals $E := \big\{ [s_1,s_2] , [t_1,t_2] \big\}$. The characteristic function of this double interval will be denoted $\chi_E$ and the gap probability written as 
$F^{(\a)}_{\mathrm P}(E;\tau_1,\tau_2) := \det(\Id - {\bf K}^{(\a)}_{\mathrm P}\chi_{E})$. In order to have readable formulas we suppressed the dependence from $ z $ in the jump matrix below. The results below are obtained in the same way as before.

\begin{problem}\label{RHPMT}
Find the sectionally analytic function $\Gamma^{(\a)}( z ) \in \mathrm{GL}(5,\C)$ on $\C/\left\{\gamma_R \sqcup \gamma_L \sqcup i\R \right\}$ such that
\be\label{RHPearceyM}
	\left\{\begin{array}{ll}
	\Gamma_+^{(\a)}( z ) = \Gamma_-^{(\a)}( z ) M^{(\a)}( z )&\\
	\nonumber&\\
	\Gamma^{(\a)}( z ) \sim \Id + \mathcal O( z ^{-1}), \quad  z  \rightarrow \infty,&\\
	\nonumber&\\
	M^{(\a)}( z ) := \left[\begin{array}{ccccc}
	1 & -{\rm e}^{\theta_{\tau_1}(s_1; z )}
	{A}
	^{-1}\chi_\gamma & {\rm e}^{\theta_{\tau_1}(s_2; z )}
	{A}
	^{-1}\chi_\gamma & -{\rm e}^{\theta_{\tau_2}(t_1; z )}
	{A}
	^{-1}\chi_\gamma & {\rm e}^{\theta_{\tau_2}(t_2; z )}
	{A}
	^{-1}\chi_\gamma
	\\
	 -{\rm e}^{-\theta_{\tau_1}(s_1; z )}
	 {A}
	 \chi_{i\R} & 1 & 0 & 0 & 0 
	 \\
	 -{\rm e}^{-\theta_{\tau_1}(s_2; z )}
	 {A}
	 \chi_{i\R} & 0 & 1 & 0 & 0 
	 \\
	  -{\rm e}^{-\theta_{\tau_2}(t_1; z )}
	  {A}
	  \chi_{i\R} & -{\rm e}^{(t_1-s_1) z +\frac{\tau_2-\tau_1}2 z ^2} & {\rm e}^{(t_1-s_2) z +\frac{\tau_2-\tau_1}2 z ^2} & 1 & 0 
	 \\
	 -{\rm e}^{-\theta_{\tau_2}(t_2; z )}
	 {A}
	 \chi_{i\R} & -{\rm e}^{(t_2-s_1) z +\frac{\tau_2-\tau_1}2 z ^2} & {\rm e}^{(t_2-s_2) z +\frac{\tau_2-\tau_1}2 z ^2} & 0 & 1 
	\end{array}\right]
	\end{array}\right.					 
\ee
\end{problem}
\bc
Let $g( z ) := \Big((\Gamma^{\emptyset})^{-1}\Big)_{1,1}( z )$ be the $(1,1)$--entry of the inverse of the solution to \ref{RHPMT} for $\a = \emptyset$. Then, for an arbitrary set of parameters $\a$,
\be
	F^{(\a)}_{\mathrm P}(E;\tau_1,\tau_2) = F_{\mathrm P}(E;\tau_1,\tau_2) \frac{\det \Big((\pa_s 
	{+}
	 a_j)^{\ell - 1} g(
	 a_j)\Big)_{\ell,j = 1}^m}{\Delta(
	 \a)},
\ee
where $F_{\mathrm P}(E;\tau_1,\tau_2)$ is the two--level gap probability for the Pearcey process with times $\tau_1,\tau_2$ and related intervals $E:= \big\{[s_1,s_2],[t_1,t_2]\big\}$.
\ec

\subsection{The multiple Hermite kernel}

As a last example, consider the multiple Hermite kernel, introduced by Br\'ezin and Hikami in \cite{BHExternalpotential} (see also \cite{Johansson1} and, for the relation with the so--called multiple Hermite polynomials, \cite{BleKui1}): 
\be\label{Hermitedet}
	K_{\mathrm H}^{(\a)} (x,y) :=  \frac{{\rm e}^{\frac{y^2-x^2}2}}{(2\pi i)^2} \oint_\gamma dz \int_{i\R +L} dw \frac{{\rm e}^{-\frac{z^2}2 - xz + \frac{w^2}2 + wy}}{z - w}\prod_{k = 1}^m\frac{w-a_k}{z-a_k}
\ee
where $\gamma$ is counter-clockwise contour encircling the points $\a := \{a_1,\ldots,a_m\}$ that are not necessarily distinct, while $L$ is a real number large enough so that the two contours do not intersect. We are interested in\footnote{As for the Pearcey case, we modified the phase of the kernel with respect to the usual notation. This implies, in particular, that the quantity we are computing is the largest particle distribution, and not the smallest one, as one could think noticing the characteristic function $\chi_{(-\infty,s]}$ (see again Remark \ref{wrongsign}).}
\be\label{GUEGP}
	F_{\mathrm H}^{(\a)}(s) := \det\Big(\Id - K_{\mathrm H}^{(\a)}\chi_{(-\infty,s]}\Big)
\ee
which is the gap probability for the $(m \times m)$ Hermitian matrix model with external potential $\mathrm{diag}(a_1,\ldots,a_m)$ (note that some of the $a$'s can be zeroes; all of them in the case without external potential).
In particular, we will give two different formulas:\\
i) The first one giving the gap probability for the Hermite matrix model with external potential in function of the one for the standard Hermite matrix model.\\
ii) The second one giving the gap probability for the Hermite matrix model (with or without external potential) as a finite--size determinant. Note that, since we already know that the Hermite kernel corresponds to a finite--rank operator, this result is not new and can be deduced without the use of Darboux transformations.\\

i) We start analyzing the following situation: suppose that $m = q + n$ and that 
\be\label{aa}
	\a = (\underbrace{0,\ldots,0}_{n},a_1,\ldots,a_q);
\ee 
i.e. there is a pole of order $n$ at $0$, and all the others $a's$ are different from zero. We denote with $F_{n,q}(s) := F_{\mathrm H}^{(\a)}(s)$ the corresponding gap probabilities given by \eqref{GUEGP} with $\a$ as in \eqref{aa}. Note that $F_n(s) := F_{n,0}(s)$ is the gap probability associated to the finite size $(= n)$ Gaussian unitary ensemble (GUE), while $F_{n,q}$ is the gap probability associated to the finite $(= m)$ GUE with external potential $A := \mathrm{diag}(0,\ldots,0,a_1,\ldots,a_q)$ (see \cite{BleKui1} and references therein). Here the Riemann--Hilbert problem we have to take into consideration is the following:
\begin{problem}\label{RHH1}
Find the sectionally analytic function $\Gamma^{(\a)}( z ) \in \mathrm{GL}(2,\C)$ on $\C/\left\{\gamma \sqcup (i\R + L)\right\}$ such that
\be\label{RHH1eq}
	\left\{\begin{array}{ll}
	\Gamma_+^{(\a)}( z ) = \Gamma_-^{(\a)}( z )\left[\begin{array}{cc}
	1 & -{\rm e}^{-\frac{ z ^2}2 - s z } z ^{-n}\ds\prod_{k = 1}^q ( z -a_k)^{-1} \chi_\gamma
	\\
	 -{\rm e}^{\frac{ z ^2}2 + s z } z ^{n}\ds\prod_{k = 1}^q ( z -a_k)\chi_{i\R+ L} & 1
	\end{array}\right]&\\
	&\\
	\Gamma^{(\a)}( z ) \sim \Id + \mathcal O( z ^{-1}), \quad  z  \rightarrow \infty.&
	\end{array}\right.					 
\ee
\end{problem}
As expected, for $q = 0$, this is the Riemann--Hilbert problem associated to a particular solution of Painlev\'e IV (the one relevant for the finite size gap probability for GUE, see for instance \cite{TWPIV}).
The following result is the analogue of the one already stated for Airy and Pearcey:
\bc
	Let $g( z ) := \Big((\Gamma^{\emptyset})^{-1}\Big)_{1,1}( z )$ be the $(1,1)$--entry of the inverse of the solution to \ref{RHH1} for $q = 0$. Then, for $\a = (\underbrace{0,\ldots,0}_{n},a_1,\ldots,a_q)$,
\be\label{BaikHermiteformula}
	F_{n,q}(s) = F_{n}(s) \frac{\det \Big((\pa_s + a_j)^{\ell - 1} g(b_j)\Big)_{\ell,j = 1}^q}{\Delta(\b)},
\ee
where $F_{n}(s)$ is the gap probability associated to the finite--size $( = n)$ Gaussian unitary ensemble.
\ec
ii) Now let's go back to the general case related to the following Riemann--Hilbert problem:
\begin{problem}\label{RHH2}
Find the sectionally analytic function $\Gamma^{(\a)}( z ) \in \mathrm{GL}(2,\C)$ on $\C/\left\{\gamma \sqcup (i\R + L)\right\}$ such that
\be\label{RHH2eq}
	\left\{\begin{array}{ll}
	\Gamma_+^{(\a)}( z ) = \Gamma_-^{(\a)}( z )\left[\begin{array}{cc}
	1 & -{\rm e}^{-\frac{ z ^2}2 - s z } \ds\prod_{k = 1}^m ( z -a_k)^{-1} \chi_\gamma
	\\
	 -{\rm e}^{\frac{ z ^2}2 + s z } \ds\prod_{k = 1}^m ( z -a_k)\chi_{i\R+ L} & 1
	\end{array}\right]&\\
	&\\
	\Gamma^{(\a)}( z ) \sim \Id + \mathcal O( z ^{-1}), \quad  z  \rightarrow \infty.&
	\end{array}\right.					 
\ee
\end{problem}
We observe the following facts:
\begin{itemize}
	\item $F_H^{\emptyset} = 1$ (it is, indeed, the gap probability for the Hermitian matrix model of size $0$).
	\item For an arbitrary $\a$, we have 
\be\label{usualformula}
	F_{\mathrm H}^{(\a)}(s) = \frac{\det\Big((\pa_s + a_j)^{\ell -1}g(a_j)\Big)_{\ell , j =1}^m}{\Delta(\a)}.
\ee
where $g( z ) := \Big((\Gamma^{\emptyset})^{-1}\Big)_{1,1}( z )$ is the $(1,1)$--entry of the inverse of the solution to \ref{RHH2} for $\a = \emptyset$.
\end{itemize}

The function $g(z)$, actually, can be computed explicitly either solving the Riemann--Hilbert problem \ref{RHH2} for $\a = \emptyset$ or either observing that, from the formula above, we have, for $m = 1$,
$$F_{\mathrm H}^{(a)}(s) = g(a) = \frac{\ds\int_{-\infty}^s \exp(-x^2/2 - ax)dx}{\ds\int_{-\infty}^\infty \exp(-x^2/2 - ax)dx}.$$

\subsection*{Acknowledgements} 

The work of M.B. is partially supported by the Natural Sciences and Engineering Research Council of Canada (NSERC) and the work of M.C. is partially supported by the ANR project DIADEMS. M.C. thanks the UMI CRM/CNRS in Montreal and the MPIM in Bonn for the hospitality and the financial support during his stay in 2013. M.C. thanks Alexander Its for some very enlightening discussions about the present work. 

\appendix
\renewcommand{\theequation}{\Alph{section}.\arabic{equation}}

\section{Integrable kernels and isomonodromic tau functions}\label{AP}
In this appendix we recall some basic facts about integrable kernels \`a la Its-Izergin-Korepin-Slavnov \cite{IIKS} and their connections with isomonodromic tau functions, recalling in particular a theorem proved in \cite{BertolaCafasso1}.
Given a piecewise smooth oriented curve ${\cal C}$ on the complex plane (possibly extending to infinity) and two 
matrix-valued functions
$$
f,g:\mathcal C\longrightarrow {\C^{k \times p}},
$$
we define the kernel $K$ as
$$
K(z,w):=\frac{f^{\mathrm T}(z)g(w)}{z-w}.
$$
We say that such kernel is \emph{integrable} if $f^{\mathrm T}( z )g( z )= 0_{p\times p}$ (so that it is non-singular on the diagonal). We are interested in the operator $K:L^2({\cal C},\C^{p})\rightarrow L^2({\cal C},\C^{p})$ acting on functions via the formula
$$
(K h)( z )=\int_{\cal C} K( z , w  ) h( w  )d w  ,
$$
and, in particular, we are interested in its Fredholm determinant $\det(\Id-K)$.
The key observation is that, denoting with $\partial$ the differentiation with respect to any auxiliary parameter on which $K$ may depend, we obtain the formula
\be
	\partial \log\det(\Id-K)=-\mathrm{Tr}((\Id+R)\partial K),
\label{resolvent}\ee
where $R$ is the resolvent operator, defined as $R=(\Id-K)^{-1}K$. Moreover $R$ is again an integrable operator, i.e.
$$
R( z , w  )=\frac{F^{\mathrm T}( z ) G( w  )}{ z - w  },
$$ and $ F, G$ can be found solving the following $k\times k$ RH problem:
\be\label{appendixRH}\left\{\begin{array}{ccc}
		 \Gamma_+( z )&=& \Gamma_-( z ) M( z ),\quad z \in{\cal C},\\
		\\
		 \Gamma( z )& =& \Id+\mathcal O( z ^{-1}),\quad z \longrightarrow\infty,\\
		\\
	 M( z )& =& \Id-2\pi i f( z ) g^{\mathrm T}( z ).
	\end{array}\right.
\ee
More precisely we have the two equalities
\bes
  F( z )= \Gamma( z ) f( z ),\qquad
		 G( z ) =( \Gamma^{-1})^{\mathrm T}( z ) g( z ).
\ees
Now suppose, as in the formula \eqref{resolvent}, that the operator $K$ (and hence the Riemann-Hilbert problem \eqref{appendixRH} depends smoothly on a certain set of parameters\footnote{In the case treated in this article, we just have one parameter, namely $s$.}. On the space of these deformation parameters, we introduce the following one-form (here below $\partial$ denotes a vector in the space of deformation parameters)
\bea
	\omega(\partial;[M])&\&:= \int_{\cal C}\mathrm{Tr}\Big( \Gamma_-^{-1}( z )\partial_ z  \Gamma_-( z )\Xi_\partial( z  )\Big)\frac{\d  z }{2\pi i},
	\label{omegamalgrange}\\
	\nonumber \Xi_\partial( z )&\& :=\partial M ( z ) M^{-1}( z )
\eea
The definition (\ref{omegamalgrange}) is posed for arbitrary jump matrices; in the case of the Riemann-Hilbert problem \eqref{appendixRH} the spontaneous question arises as to whether $\omega_M$ in (\ref{omegamalgrange}) and the Fredholm determinant are related. The answer is positive  within a certain explicit correction term, as in the theorem below

\begin{theorem}[\cite{BertolaCafasso1}]\label{thdetMalgrange}

Let $f( z ;\vec s), g( z ;\vec s): \mathcal{C} \times S \longrightarrow {\mathbb{C}^k}$ and consider the Riemann-Hilbert problem with jumps as in \eqref{appendixRH}. Given any vector field $\partial$ in the space of the parameters $S$ of the integrable kernel we have the equality
\bes
		\omega(\partial;[M])=\partial\log\det(\Id-K)+H(M),
\ees
	where $\omega(\partial;[M])$ is as in (\ref{omegamalgrange}) and
$$
H(M):=H_1(M)-H_2(M)=\int_{\cal C} \mathrm{Tr}\Big(\partial f'^{\mathrm T} g+f'^{\mathrm T}\partial g \Big)d z  -2\pi i\int_{\cal C}\mathrm{Tr}( g^{\mathrm T} f'
\partial g^{\mathrm T} f) d z .
$$
\end{theorem}
In the cases we treat in the article, moreover, we have $H(M)=0$.
Hence it is possible to define, up to normalization, the isomonodromic tau function $\tau_{JMU}:=\exp(\int\omega)$ and this object, thanks to the previous theorem, will coincide with the Fredholm determinant $\det(\Id-K)$ (up to constant terms). One more interesting formula arises in the case when, choosing appropriately a diagonal matrix $T( z )$ and defining $\Psi( z ) := \Gamma( z ){\rm e}^{T( z )}$, this latter satisfies a Riemann--Hilbert problem with constant jumps. Then, as it is proven in \cite{BertolaIsoTau}, we have the following equality
\be\label{appisomtau}
	\int_{\cal C}\mathrm{Tr}\Big( \Gamma_-^{-1}( z )\partial_ z  \Gamma_-( z )\Xi_\partial( z  )\Big)\frac{\d  z }{2\pi i} = -\res_{ z  = \infty}\mathrm{Tr}\left(\Gamma^{-1}\left(\partial_ z  \Gamma\right)\partial_s T\right)
\ee
giving directly the log--derivative of the isomonodromic tau function in terms of the entries of the asymptotic expansion of $\Gamma$ at infinity.

\section{Proof of  Theorem \ref{maintheorem}}\label{appMarco}
\subsection{Schlesinger tranformations}
\label{secSchles}
We start with a more general problem than the one addressed in Theorem \ref{maintheorem}: the problem has a distinguished history, having been addressed systematically in \cite{JMU2} but for the case of differential equations with rational coefficients, rather than for general Riemann Hilbert problems. 
 
The approach we take is strictly more general: the proof of the theorem shall also be entirely different than the one contained in said paper. Let us mention that the proof in \cite{JMU2} (contained in section 3.4 of that paper) is a proof by induction on the number of ``elementary'' steps and with many details left to the good will of the reader. We propose a direct (not inductive) proof.
In the applications we are aiming at, the matrix $\Gamma(z)$ is the solution of some RHP and hence it has some additional properties; however the notion of Schlesinger transformation does not require any of that. In fact the only data that need to be supplied are germs of (formal) analytic functions near a finite collection of points.
 
\bd
\label{defSchles} Let $\mathcal A, \mathcal B$ be two finite collection of points in $\C\mathbb P^1$ (not necessarily disjoint).\\
Let $\Gamma(z)\in Mat(N\times N,\C)$ 
be a (formal) germ of analytic function at each point of $\mathcal C := \mathcal A \cup \mathcal B$ and also near $z=\infty$ (which may already belong to $\mathcal C$)  such that $\Gamma^{-1}(z)$ is also a (formal) germ of analytic function, and $\Gamma(z) = \1 + \mathcal O(z^{-1})$ near $z=\infty$.   For each point $b\in \mathcal B$ let $ K_b = {\rm diag}(k_{b,1}, \dots, k_{b,n})$ with nonnegative integers and similarly for each point $a\in \mathcal A$ let $ L_a = {\rm diag}(\ell_{a,1}, \dots, \ell_{a,n})$ with nonnegative integers satisfying the  ``zero index'' condition 
\be
\label{zeroindex}
\sum_{b\in \mathcal B} \ \tr  K_b = \sum_{a\in \mathcal A} \tr  L_a \ .
\ee
We also require that if $c\in \mathcal A\cap \mathcal B$ then $ K_c L_c = 0$, i.e., they have nonzero entries in different places along the diagonal. We shall set $K_c ={\rm diag}(0,\dots, 0)$ and/or $ L_c={\rm diag}(0,\dots, 0)$ if $c$ does not belong to $\mathcal B$, $\mathcal A$, respectively.
The {\bf Schlesinger transform of $\Gamma$} subordinated to these data is a rational matrix $R(z) = R_{
{\le\{{\mathcal A\, \mathcal B}\atop{ L\,K}\ri\}}}(z)$  such that 
\begin{itemize}
\item it is analytic and analytically invertible in $\C\setminus \mathcal C$;
\item The matrix $R(z)  \Gamma(z) z_c^{K_c-L_c}$ is a  (formal) analytic series at $\mathcal C\cup \{\infty\}$, with (formal) analytic inverse;
\item $R(z)  \Gamma(z) z_c^{K_c-L_c} = \1 + \mathcal O(z^{-1})$ near $z=\infty$. 
\end{itemize}
\ed
\bp
\label{propR} The following statements hold:\\
{\bf [1]} If $R(z) = R_{
{\le\{{\mathcal A\, \mathcal B}\atop{ L\,K}\ri\}}}(z)$ exists, it is unique;\\
{\bf [2]} The function $R(z)$ has poles only at the points in $\mathcal B$;\\
{\bf [3]} The function $R^{-1}(z)$ has poles only at the points in $\mathcal A$.\\
{\bf [4]} The determinant satisfies
\be
\det R(z) = \prod_{\mu=1}^n 
\frac
{ \prod_{a\in \mathcal A\atop a\neq\infty} (z-a)^{\ell_{a,\nu}}}
{ \prod_{b\in \mathcal B\atop b\neq\infty} (z-b)^{k_{b,\nu}}}
\label{detR}
\ee
\ep
{\bf Proof}. 
{\bf [1]} Suppose $\wt R$ satisfies the same defining conditions: then $\wt R(z) R^{-1}(z)$ is {\em a priori} a rational function with poles only at $\mathcal C$. On the other hand it cannot have any singularity at $\mathcal C$ because $\wt R(z) R^{-1}(z) = \le(\wt R(z) \Gamma(z) z_c^{K_c-L_c}\ri)\le( z_c^{L_c-K_c} \Gamma^{-1}(z) R^{-1}(z)\ri)$ and each factor in the brackets is analytic by assumption (at first formally, but since the rhs must have a convergent Laurent expansion, also non-formally). At $z=\infty$ we must have also  $\wt R(z) R^{-1}(z) \to \1$ and thus by Liouville's theorem $\wt R(z) R^{-1}(z) \equiv 1$. \\
{\bf [2] , [3]}. Near $c\in \mathcal C$ we have $R(z) = \mathcal O(1) z^{L_c-K_c} \Gamma^{-1}(z)$ and hence the Laurent tail near $z=c$ is bounded, and the only poles can be at $c\in \mathcal B$  where $K_b$ is nonzero. Similarly for $R^{-1}$. \\
{\bf [4]} The determinant of $R(z)$ is a rational function with zeroes and poles at $z\in \mathcal C$ of the indicated orders. The normalization at infinity fixes the overall constant. \QED

\br[Group property]
Schlesinger transformations have a group property as follows: given $\Gamma, R_{
{\le\{{\mathcal A\, \mathcal B}\atop{ L\,K}\ri\}}}$ as described, then 
\be
\Gamma_{
{\le\{{\mathcal A\, \mathcal B}\atop{ L\,K}\ri\}}}(z):= R_{
{\le\{{\mathcal A\, \mathcal B}\atop{ L\,K}\ri\}}} (z)
\Gamma (z)z_c^{K_c-L_c} \label{gammasch}
\ee
defines a new germ of analytic series\footnote{If we allow $K_c=0=L_c$ then this simply means that this particular point is ``not involved'' in a transformation but it may be involved in a subsequent one.} which, in turn, can be used as a starting point for an additional Schlesinger transformation. Then the reader is invited to verify that the result of two such subsequent transformation with indices $(K, L), (\wt K, \wt L)$ is the same as the one with indices $(K + \wt K, L + \wt L)$. 
\er

\subsection{Variation of the Malgrange differential under Schlesinger transformations}
Given the set of points $\mathcal C$ for the Schlesinger transformation we consider the union of small, nonintersecting disks $\D := \bigsqcup _{c\in \mathcal C} \D_c$.
The data of the degrees $K, L$ can be encoded in a single, piecewise diagonal meromorphic matrix  function
\bea
D(z) = \le\{ \begin{array}{cc}
z_c^{(L_c -K_c)}\ &  z\in \D_c\\
1 & z\in\C \setminus \D
\end{array}
\ri.\ \ \Rightarrow \ \ 
D(z) = \prod_{c\in \mathcal C} z_c^{(L_c -K_c)\chi_{\D_c}}
\label{Ddef} 
\eea
Later on we shall also consider deformations with respect the positions of the points $c\in \mathcal C$; we immediately stipulate that an expression  involving derivatives w.r.t. $c\in \mathcal C$ do leave the position of the disks fixed. Thus, for example, 
\be
D^{-1}(z) \pa_{c_0} D(z) := \frac {K_{c_0}-L_{c_0}}{z-c_0} \chi_{\D_{c_0}}.\label{dc0}
\ee
For the remainder of this section, we return to the original setting, whereby $\Gamma(z)$ is the solution of some RHP \eqref{RHPGamma} and is used as a germ of analytic function at the points involved in the Schlesinger transformation.  We can then reformulate the Schlesinger transformation  as a RHP for a new matrix 
$\wt \Gamma(z)$ with jumps on $\Sigma \cup  \pa \D$   and the properties
\be
\label{RHPtildeGamma}
\wt \Gamma_+ = \wt \Gamma_- M\ ,\ \ \ z\in \Sigma;\ \ 
\wt \Gamma_+(z) = \wt \Gamma_-(z) z_c^{K_c-L_c} ,\ \ z\in \pa \D_c;\ \ 
\wt \Gamma(z) = \1 + \mathcal O(z^{-1}) \ ,\ \ \ z\to\infty.
\ee
\bl
\label{wtgammagamma}
The relationship between the solution $\wt \Gamma(z)$  of  the RHP \eqref{RHPtildeGamma} and $\Gamma(z)$ is  
\bea
\wt \Gamma(z) &=& R_{
{\le\{{\mathcal A\, \mathcal B}\atop{ L\,K}\ri\}}}(z) \Gamma(z)\ ,\ \ z \in \C \mathbb P^1 \setminus \D,\\
\wt \Gamma(z) &=& R_{
{\le\{{\mathcal A\, \mathcal B}\atop{ L\,K}\ri\}}}(z) \Gamma(z)z_c^{K_c-L_c} \ ,\ \ z \in  \D\ .
\eea
\el
{\bf Proof.}
It follows easily that $R(z):= \wt \Gamma(z)  \Gamma^{-1}(z)$ in $\C \setminus \bigcup_{c\in\mathcal C} \D_c$ is a function without jumps on $\Sigma$ that extends to a rational function with poles at $\mathcal B$ and with inverse with poles at $\mathcal A$: namely it is precisely the matrix $R$ of Definition \ref{defSchles}. 
On the other hand, inside each disk $\D_c$, the matrix $\wt \Gamma$ is simply $R\Gamma z_c^{K_c-L_c}$, namely, the matrix $\Gamma_{
{\le\{{\mathcal A\, \mathcal B}\atop{ L\,K}\ri\}}}(z)$ of \eqref{gammasch}. \QED

Correspondingly we shall consider 
\bea
\label{b7}
\Omega\le(\pa; [M],
{\le\{{\mathcal A\, \mathcal B}\atop{ L\,K}\ri\}}\ri) := \frac 12 
\int_{\Sigma }  \frac {\d z}{2i\pi} \tr \bigg(\wt \Gamma_-^{-1} \wt \Gamma_-' \pa M M^{-1} + \wt \Gamma_+^{-1} \wt \Gamma_+' M^{-1} \pa M  
\bigg)  +\nonumber\\+
 \frac 1 2 \sum_{c\in \mathcal A\cup \mathcal B\atop} \oint_{\pa \D_c} \frac{\d z}{2i\pi}
 \tr \le(\bigg(\wt \Gamma_-^{-1} \wt \Gamma_-'  + \wt \Gamma_+^{-1} \wt \Gamma_+' 
\bigg) \pa (z_c)^{
{L}_c -
{K}_c}(z_c)^{
{K}_c -
{L}_c}\ri). \label{b8}
\eea
The reader is invited to verify that the second term in the last integral in  \eqref{b8} yields the same contribution as the first (using that $\wt \Gamma_+ = \wt \Gamma_- z_c^{K_c-L_c}$ on $\pa \D_c$ and seeing that the additional term coming from the second one above, has at most a double pole without residue) and thus we can write a shorter equivalent expression 
\bea
\Omega\le(\pa; [M],
{\le\{{\mathcal A\, \mathcal B}\atop{ L\,K}\ri\}}\ri) := \frac 12 
\int_{\Sigma }  \frac {\d z}{2i\pi} \tr \bigg(\wt \Gamma_-^{-1} \wt \Gamma_-' \pa M M^{-1} + \wt \Gamma_+^{-1} \wt \Gamma_+' M^{-1} \pa M  
\bigg)  +\nonumber\\+
 \sum_{c\in \mathcal A\cup \mathcal B\atop} \oint_{\pa \D_c} \frac{\d z}{2i\pi}
 \tr \bigg(\wt \Gamma_-^{-1} \wt \Gamma_-' \pa D D^{-1}
\bigg).
\eea
\bp
\label{propSchles}
Under any deformation $\pa$ of $M = M(z;\un t)$,  or of the position of poles/zeroes of any point $c\in \mathcal A\sqcup B$\footnote{For the purpose of deformations, we allow poles/zeroes to move independently, even if they appear as both (i.e. a zero in some column, and a pole in another, at the same point $c$.)} we have 
\bea
\Omega\le(\pa; [M],
{\le\{{\mathcal A\, \mathcal B}\atop{ L\,K}\ri\}}\ri)  - \Omega\le(\pa; [M]\ri) =  
\sum_{c\in \mathcal A\cup \mathcal B} \res_{z=c} \tr \le(R^{-1} R' \pa \S \S^{-1}\ri) \d z +\oint_{\pa \D} \frac{\d z}{2i\pi}
 \tr \bigg(\Gamma^{-1}  \Gamma' \pa D D^{-1}
\bigg)\\
\S(z):= \Gamma(z;\un t) D(z).
\eea
where 
$R(z) = R_{
{\le\{{\mathcal A\, \mathcal B}\atop{ L\,K}\ri\}}}(z)$ is the rational matrix of the Schlesinger transformation in Def. \ref{defSchles}. 
\ep
{\bf Proof.}
Recall that (Lemma \ref{wtgammagamma}) in the exterior of $\D$ we have  $\wt \Gamma(z)  = R(z) \Gamma(z)$ and hence
\bea
&\& \frac 1 2 \int_{\Sigma }\frac {\d z}{2i\pi}\tr \bigg(\wt \Gamma_-^{-1} \wt \Gamma_-' \pa M M^{-1} + \wt \Gamma_+^{-1} \wt \Gamma_+' M^{-1} \pa M  
\bigg)  = \nonumber \\
&\& = \frac 1  2\int_{\Sigma }\frac {\d z}{2i\pi}\tr 
\bigg(
  \Gamma_-^{-1}  \Gamma_-' \pa M M^{-1} + \Gamma_+^{-1}  \Gamma_+' M^{-1} \pa M  +
  R^{-1} R'  \Gamma_- \pa M M^{-1}\Gamma^{-1}_-   + R^{-1} R'  \Gamma_+M^{-1}  \pa M\Gamma^{-1}_+ \bigg) =\nonumber \\
 &\&= \Omega(\pa; [M])  +\frac 1 2   \int_{\Sigma }\frac {\d z}{2i\pi}\tr 
\bigg(
  R^{-1} R'  \Gamma_- \pa M M^{-1}\Gamma^{-1}_-   + R^{-1} R'  \Gamma_+M^{-1}  \pa M\Gamma^{-1}_+ \bigg).\label{213}
\eea
Now we use the easily verified identity 
$$
 \Gamma_- \pa M M^{-1}\Gamma^{-1}_- + \Gamma_+M^{-1}  \pa M\Gamma^{-1}_+ = 
 2 \Delta_\Sigma \le( \pa \Gamma \Gamma^{-1}\ri),
$$
where $\Delta_\Sigma (F)(z) = F_+(z) - F_-(z),\ z\in \Sigma$. 
Then the last term in  \eqref{213} is  reduced (by Cauchy's theorem) to an integral  on $\pa \D = \bigcup _c \pa \D_c$
\bea
  \int_{\Sigma }\frac {\d z}{2i\pi}\tr 
\bigg(
  R^{-1} R' \Delta_{\Sigma}(\pa \Gamma \Gamma^{-1})\bigg) = \int_{\pa \D} \frac {\d z}{2i\pi}\tr 
\bigg(
  R^{-1} R' \pa \Gamma \Gamma^{-1}\bigg). \nonumber
\eea
Using the cyclicity of the trace and recalling the definition \eqref{Ddef} of $D(z)$ together with the fact that $\wt \Gamma_- = R \Gamma$ for $z\in \pa \D$ (while $\Gamma$ has no jump on $\pa \D$) , we have 
\bea
\Omega\le(\pa; [M],
{\le\{{\mathcal A\, \mathcal B}\atop{ L\,K}\ri\}}\ri)  - \Omega\le(\pa; [M]\ri) =  \int_{\pa \D} \frac {\d z}{2i\pi}\tr 
\bigg(
  R^{-1} R' \pa \Gamma \Gamma^{-1}\bigg)  +\nonumber\\
  +\oint_{\pa \D} \frac{\d z}{2i\pi}
 \tr \bigg(R^{-1} R'       \Gamma \pa D D^{-1} \Gamma^{-1}+   \Gamma^{-1}  \Gamma' \pa D D^{-1}
\bigg)\nonumber
\eea
and the first two terms above combine to form the first term in the statement.
{\bf Q.E.D.}

The goal of the next section (which is the most important) is to show that the right hand side in Prop. \ref{propSchles} is the logarithmic variation of the finite determinant of the characteristic matrix in Def. \eqref{defGmatrix}.
This is achieved in Cor. \ref{cor22}.
\subsection{The determinant of the characteristic matrix and its deformations}
Associated to the  region $\D$  is the Hilbert space $\H= L^2(\pa \D, |\d z|)$, which, by virtue of the form of $\D = \bigsqcup_{a\in \mathcal A} \D_a \cup \bigsqcup_{b\in \mathcal B} \D_b$, is the direct sums of several copies of $L^2(S^1)$. We also have the Cauchy boundary operators 
\be
C_\pm: \H\to \H_\pm,  \ \ \ C_\pm[f](z) = \oint_{\pa \D} \frac {f(\z) \d \z}{(\z-z_\pm)2i\pi}
\ee
which projects onto the space $\H_+$ of analytic functions on $\D$.
\footnote{
Remember that $\D$ is disconnected, so an analytic function on $\D$  is really collection of analytic functions on each of the $\D_c$'s.
Quite opposite instead, since $\D_-$ is connected, an analytic function in $\H_-$ is a ``single'' analytic function.}
\br[Important but simple]
\label{reminfty} If $\infty \in \D_-$ then none of the disks $\D_c$ is unbounded: then  $H_+$ consists of analytic functions in each of the (finite) disks and $\H_-$ consists of analytic functions on the complement {\em that tend to zero at infinity}. 
Thus the constant functions belong to $\H_+$. 

Conversely, if $\infty \in \D_+$ then one of the disks $\D_c$ is a disk at infinity; then $H_+$ consists of analytic functions that {\em vanish at $z=\infty$} and $\H_-$ of all analytic functions in complement $\C \setminus \D$. In particular the constant functions belong to $\H_-$, contrary to the previous case.
\er

{\bf Notation:}
In the sequel we shall use $\H, \H_\pm$ to denote also the space of (row) vector functions or matrix-valued functions with entries in $\H, \H_\pm$; the meaning (scalar or vector or matrix) should be clear within the context.

Consider the finite dimensional vector subspaces of $\H_-$
\bea
V:= C_- [\H_+ \S^{-1}]\ ,\ \ W:= C_-[\H_+ \S]\ ,\ \ \H_+  = \{\hbox{ row vectors of analytic functions in $\D_+$}\} \nonumber
\\
 \S(z):= \Gamma(z) D(z)
\label{spacesVW}
\eea
The finite dimensionality is clear because in each disk $\D_C$ we have  $\S(z) = \Gamma(z) z_c^{L_c-K_c}$. Thus both $\S(z)$  and $\S(z)^{-1}$ have at most finite-order poles and either the points $b\in \mathcal B$ or $a\in \mathcal A$ (respectively).

\br[Important]
\label{formalrem}
If $\Gamma$ is only a formal germ of analytic function, then the Cauchy projectors shall be understood as the projector onto the Laurent tail, and in that case $C_+ := Id + C_-$.  We shall nevertheless continue using it as an integral operator for clarity. 

For example, in all the cases discussed in Section \ref{sectGP}, the contour $\Sigma$ extends to $\infty$ but  nonetheless $\Gamma(z)$ admits a formal expansion at $z=\infty$ {\em independent} of the sector. Therefore $\Gamma$ does define a formal germ of analytic function at $\infty$. 
\er
\bp
\label{invertG}
The  matrix $R(z) = R_{\le\{
{\mathcal A\,\mathcal B}\atop{L\,K}
\ri\}}$ exists (and thus is unique) if and only if $\mathcal G: V\to W$ (defined in \eqref{spacesVW})  given by $
\mathcal G(v)= C_-[v \S]$ 
is invertible and the inverse is 
\be
\mathcal G^{-1}(w) = C_-\le[w \S^{-1} R^{-1}\ri] R:W \to V 
\label{inverseG}.
\ee
\ep
{\bf Proof.} This is a classical result that we prove for the interest of the reader.
First of all we need to verify that  $\mathcal G$ indeed maps to $W$; to this end 
let $v = C_-[h_+  \S ^{-1}]\in V $ then 
\bea
C_-[v  \S ] =  C_+ [v  \S ] -v  \S  = C_+[v \S ] -C_-[h_+ \S ^{-1}] \S  =
\overbrace{C_+[C_-[h_+ \S ^{-1}] \S ] + h_+} ^{\in \H_+} - C_+[h_+ \S ^{-1}] \S.  \nonumber
\eea 
Taking the $C_-$ projection of both sides (and recalling that $C_-\circ C_- = -C_-$) leaves the left side invariant and thus 
$$
C_-[v  \S ] = C_-[C_+[h_+ \S ^{-1}] \S  ]\in C_-[\H_+  \S ] = W.
$$
This proves the well posedness of $\mathcal G$. As for the claims proper:\\{}
$[\Leftarrow]$ Suppose that $\mathcal G:V\to W$ is invertible and let $L(z)$ the matrix $\mathcal G^{-1}[C_-[\S]]$ (meaning, the result of adjoining all the rows of $\mathcal G^{-1}[C_-[{\bf e}_j^t\S]]$).
Then we claim that $R(z) = \1 - L(z)$ is the solution and $R_+(z)  = R(z) \S$ is a (formal) germ of analytic function in $\D_+$. To see it we need to verify that $R_+\in \mathcal H_+$ (in the sense of matrices with entries in $\H_+$). But this is clear 
$$
C_-[(\1 - L)\S] = C_-[\S] - C_-[L\S] = C_-[\S] - C_-[\S] =0  
$$
since by assumption $\mathcal G (L)= C_-[L\S] = C_-[\S]\in V$ (row-wisely). We remark (leaving the verification to the reader) that the above argument does not rely in the least on $\infty$ being on the ``exterior'' of $\D$. The only difference between the two cases is that the constant functions belong to $\H_+$ if $\infty$ is in  $\D_-$ and to $\H_-$ if $\infty\in \D_+$.   We, however,  still need to show that both $R, R_+ $ are analytically invertible. 

Note that, within $\D_c$,  $\det R_+  = \det R \det \S =\det R \det \Gamma z^{\tr (L_c-K_c)}$ and since $\Gamma$ is (formally) analytic in $\D_+$ and thanks to the zero index condition \eqref{zeroindex} then 
the number of zeroes (with multiplicities) of $\det R_+$ in $\D_+$ is the same as the number of zeroes of $\det R$ in $\D_-$; for this reason it is sufficient to show that $\det R(z)\neq 0$ for $z\in \D_-$.  
Suppose $\det R(z_0)=0$ at some point $z_0\in \D_-$. Then there is a nonzero constant row-vector ${\bf f}$ such that ${\bf f} R(z)$ vanishes at $z_0$. Thus $
\phi_-(z):= \frac{{\bf f}}{z-z_0} R (z)
$
is analytic at $z=z_0$ and thus belongs to  $\H_-$.  Define $\phi_+(z):= \frac{{\bf f}}{z-z_0} R_+(z) \in \H_+$ (the point $z_0\not \in \D_+$). Then 
$$
\phi_+(z) = \phi_-(z) \S(z) \ ,\ \ z\in \pa \D\qquad 
\lim_{z\to\infty} \phi(z) = \vec 0.
$$
Note that the last statement is valid either if $\infty\in \D_\pm$.  We show now that $w  =- \phi_- =  C_-[ \phi_+ \S^{-1}] \in W$ is in the kernel of $\mathcal G^{-1}$ thus contradicting the assumed invertibility. Indeed :
\bea
\mathcal G^{-1} \le[C_-[ \phi_+\S^{-1}]\ri] = 
\mathcal G^{-1} \le[C_-[ \phi_-]\ri] = 
-\mathcal G^{-1} \le[\phi_-\ri] = -C_-\le[ 
 \frac{{\bf f}}{z-z_0} R(z)\S
\ri]  = -C_-\le[ 
 \frac{{\bf f}}{z-z_0} R_+(z)
\ri] = 0 \nonumber
\eea 
where we have used in the last step that $z_0\in \D_-$ and hence $\frac{{\bf f}}{z-z_0} R_+(z)\in \H_+$ (and $C_-[\H_+]=0$). This concludes the proof of the sufficiency.   Note that the uniqueness follows now from Prop. \ref{propR}.\\
 $[\Rightarrow]$ Now suppose that the matrix $R=R_{\le\{
{\mathcal B\,\mathcal A}\atop{K\,L}\ri\}}$ exists: we claim that the inverse is given by \eqref{inverseG}.
 Indeed, using $C_+-C_- = \Id$ and setting $R_+ :=R \S\in \H_+$ we have\footnote{The statement $R_+ = R \S \in \H_+$ is by definition of $R$, second bullet in Def. \ref{defSchles}.} 
\bea
\mathcal G[C_-[wR_+^{-1}]R] = C_-\le[C_-[ wR_+^{-1}]R\S\ri] =
C_-\le[C_-[wR_+^{-1}]R_+\ri] =C_-\bigg[-wR_+^{-1}R_+  +\underbrace{ C_+ [wR_+^{-1}]R_+ }_{\in \H_+} \bigg] =  w,\nonumber\\ 
\mathcal G^{-1} \mathcal G[v] = C_-[ C_-[v\S]R_+^{-1}]R = C_-[ -v\S R_+^{-1} + \cancel{C_+[v \S]R_+^{-1}}]R =
 - C_-\big[-\underbrace{v R^{-1}}_{\in \H_-}\big] R = v R^{-1} R = v.\nonumber
\eea
Thus $\mathcal G$ is inverse and the inverse is written by \eqref{inverseG}.\QED

As we shall promptly see, the matrix representation of $\mathcal G:V\to W$ is precisely the characteristic matrix in Def. \ref{defGmatrix}: to this end  we  fix appropriate bases in $V, W$:
\be
V = \hspace{-12pt}\bigoplus_{a\in \mathcal A, \mu\leq m;\atop 1\leq  \ell \leq \ell_{a,\mu}} \hspace{-12pt}\C \{v_{a,\mu,\ell}\}\ ,\qquad
v_{a,\mu,\ell} = 
 C_-\big[ {\bf e}_\mu^t z_a^{\ell_{a,\mu}- \ell}\overbrace{ D^{-1}(z) \Gamma^{-1}(z)}^{=\S^{-1}(z) }\big]  =  C_-\le[ \frac{{\bf e}_\mu^t  \Gamma^{-1}(z)}{(z_a)^{\ell- \delta_{a\infty}}} \ri] ,
 \label{Vbasis}
\ee
\be
W = \hspace{-12pt}\bigoplus_{b\in \mathcal B, \nu\leq m;\atop 1\leq  k \leq k_{b,\nu}} \hspace{-10pt}\C \{w_{b,\nu,k}\}\ ,\qquad
w_{b,\mu,k} = 
 C_-\big[ {\bf e}_\nu^t z_b^{k_{b\nu}- k}\Gamma^{-1}(z) \overbrace{\Gamma (z)D(z)}^{=\S(z)}\big]  =  C_-\le[ \frac{{\bf e}_\nu^t}{(z_b)^{k- \delta_{b\infty}}} \ri].
 \label{Wbasis}
\ee
The reasons for the shifts $\delta_{a\infty}, \delta_{b\infty}$ in \eqref{Vbasis}, \eqref{Wbasis} is the fact that if $\infty\in \D_+$ (i.e. it belongs to $\mathcal A\cup \mathcal B$) then the constant functions belong to $\H_-$ (see Rem. \ref{reminfty}).
\bp
\label{characteristicmatrix}
The matrix $G$ representing $\mathcal G = C_-[\bullet  \S ]:V\to W$ in the bases \eqref{Vbasis}, \eqref{Wbasis}  is the characteristic matrix in Def. \ref{defGmatrix}:
\be
G_{(b,\nu,k); (a,\mu,\ell)} =  \res_{z=a}\res_{\z=b}\frac { {\bf e}_\mu^t \Gamma^{-1}(z) \Gamma(\z) {\bf e}_\nu   \d z
\d \z
}{(z_a)^{\ell- \delta_{a\infty}} (z-\z) (\z_b)^{k_{b,\nu}+1-k- \delta_{b\infty}}}.
\ee
\ep
{\bf Proof.}
The proof is a straightforward exercise using the  eq. \eqref{componentW}  of Lemma \ref{lemmaW} (immediately hereafter) and using that $D(z) = z_b^{-K_b}$ when $z\in \D_b$.  {\bf Q.E.D.}
\bl
\label{lemmaW}
{\bf [1]} Let $w\in W = C_-[\H_+  \S ]$; then its components in the basis \eqref{Wbasis} are given by 
\be
 \label{componentW}
 \res_{z_b=0} w(z)\cdot  {\bf e}_\nu z_b^{k-\delta_{b\infty}} \frac{\d z_b}{z_b}\ ,\ \  1 \leq k \leq k_{b,\nu}.
 \ee
 {\bf [2]}  Let $v\in V = C_-[\H_+  \S ^{-1}]$; then its components in the basis \eqref{Vbasis} are given by 
 \be
 \label{componentV}
 \res_{z_a=0} v(z)\cdot \Gamma(z)  {\bf e}_\mu z_a^{\ell - \delta_{a\infty}} \frac{\d z_a}{z_a}\ ,\ \ 1\leq \ell \leq \ell_{a,\mu}.
 \ee
\el
{\bf Proof.} It is sufficient to verify 
$$
\res_{z=b} w_{b',\nu',k'}(z)\cdot  {\bf e}_\nu z_b^{k-\delta_{b\infty}} \frac{\d z_b}{z_b} = \delta_{bb'}\delta_{kk'} \delta_{\nu'\nu}
 \ \ \ \ \ 
  \res_{z=a} v_{a',\mu',\ell'} (z)\cdot \Gamma(z)  {\bf e}_\mu z_a^{\ell - \delta_{a\infty}} \frac{\d z_a}{z_a}=
\delta_{aa'}\delta_{\ell\ell'} \delta_{\mu'\mu}.
$$
The verification of the first is absolutely straightforward and left to the reader. To verify the second we observe that the quantity to be computed is the one below 
$$
  \res_{z=a}\res_{\z=a'} \frac{{\bf e}_{\mu'}^t  \Gamma^{-1}(\z)}{{\z_{a'}}^{\ell' - \delta_{a'\infty}} (\z-z)} \Gamma(z)  {\bf e}_\mu z_a^{\ell - \delta_{a\infty}} \frac{\d z_a}{z_a}.
$$
Clearly, this is zero whenever $a\neq a'$ because the result of the residue in $\z=a'$ is analytic at $z=a$ (and vanishes  at $a$ if $a=\infty$).
If $a=a'$ 
$$
  \res_{z=a}\res_{\z=a} \frac{{\bf e}_{\mu'}^t  \Gamma^{-1}(\z)}{{\z_{a}}^{\ell'-\delta_{a\infty}} (\z-z)} \Gamma(z)  {\bf e}_\mu z_a^{\ell-\delta_{a\infty}} \frac{\d z_a}{z_a} =  
   \res_{z=a}\res_{\z=a} {\bf e}_{\mu'}^t \frac{( \Gamma^{-1}(\z)\Gamma(z)-\1)}{{\z_{a}}^{\ell'-\delta_{a\infty}} (\z-z)}   {\bf e}_\mu z_a^{\ell-\delta_{a\infty}} \frac{\d z_a}{z_a}  + \res_{z=a}\res_{\z=a} \frac{ {\bf e}_{\nu'}^t{\bf e}_\mu  z_a^{\ell-\delta_{a\infty}} }{{\z_{a}}^{\ell'-\delta_{a\infty}} (\z-z)}   \frac{\d z_a}{z_a}. 
$$
The first term gives zero because the result of the residue in $\z$ is an analytic function in $z$ (and vanishing at $a=\infty$ if $a=\infty$) and $\ell \geq 1$, and the second term clearly yields $\delta_{\mu'\mu}\delta_{\ell\ell'}$. 
{\bf Q.E.D.}

\subsubsection {Proof of Thm. \ref{maintheorem1}.}
\label{proofthm1}
The proof is simply the collection of Prop. \ref{invertG} (stating that the $R(z)$ exists iff $\mathcal G$ is invertible), together with the fact that the characteristic matrix $G_{
{\le\{{\mathcal A\, \mathcal B}\atop{ L\,K}\ri\}}}$ is nothing else but the matrix representation of $\mathcal G$ in a special basis (Prop. \ref{characteristicmatrix}). \QED
\subsubsection*{Variation of $\det G$}
We now assume that $\Gamma(z)$ is allowed to vary smoothly (while remaining in (formal) $\H_+$ of the region $\D$, of course) with respect to some additional manifold of parameters , which we shall denote generically by $\un t$. Then, clearly, the matrix $G$ representing $\mathcal G$ in the bases  \eqref{Vbasis}, \eqref{Wbasis} also inherits a dependence on these $\un t$. In addition to that, the matrix $G$ depends on the positions of the points $\mathcal A \cup \mathcal B$: we shall denote this dependence generically with $\un c$. 

We want to express the logarithmic derivative of $\det G(\un t,\un c)$ with respect any deformation (of $\Gamma$ or of the positions $\un c$) in terms of an integral. 
Before proceeding we consider the general
\bl
\label{lemmamoving}
Let $\mathcal G = C_-[\bullet \S]:\H_- \to \H_-$ (in particular mapping $V\to W$)  and fix bases\footnote{Or any other bases depending smoothly on the parameters.}\eqref{Vbasis}, \eqref{Wbasis} $\{v_j\}, \{w_s\}$ (we use generic indices $j, s$ to label elements of the basis) in $V, W$. Let $\pa$ denote the derivative with respect to any of the deformation parameters the bases depend on;  then the following maps define endomorphisms of $V, W$;
\bea
\mathcal V_\pa[v_j] = (\mathcal G^{-1}\pa \mathcal G) v_j +  \pa  v_j  - \sum_{k}G_{kj}\mathcal G^{-1} \pa w_k,
\label{Vmove}\\
\mathcal W_\pa[w_s] =  \pa \mathcal G \mathcal G^{-1} w_s-  \pa w_s+ \mathcal G  \sum_j G^{-1}_{js}\pa v_j.
\label{Wmove}
\eea
Moreover, if $[G_{ij}]$ is the matrix representative of $\mathcal G:V \to W$, then  
\be
\pa \ln \det G = \tr_V \mathcal V_\pa = \tr _W \mathcal W_\pa.
\ee
\el
{\bf Proof}. Differentiating the relation $\mathcal G v_j =\sum_k G_{kj} w_k$ w.r.t. $\pa $ we obtain (we denote by a dot $\dot{}$ the derivative $\pa$)
\bea
\dot \mathcal G v_j + \mathcal G \dot v_j &\& = \sum_{k=1}^m \le(\dot G_{kj} w_k +  G_{kj} \dot w_k \ri).
\label{dotG}
\eea
If we apply $\mathcal G^{-1}$ to both sides of \eqref{dotG} we find 
\bea 
(\mathcal G^{-1}\dot \mathcal G) v_j +  \dot  v_j &\& =\sum_{k=1}^m \le(\dot G_{kj} \mathcal G^{-1} w_k +  G_{kj} \mathcal G^{-1}\dot w_k\ri)= \sum_{k=1}^m \le(\dot G_{kj} \sum_{\ell=1}^m G^{-1}_{\ell k} v_\ell +  G_{kj}\mathcal G^{-1} \dot  w_k \ri).
\label{movingbases}
\eea
Rearranging the terms in \eqref{movingbases} and tracing over $V$ yields the assertion $\pa \ln \det G = \tr_V\mathcal V_\pa$. 
For the other case, we multiply by $G^{-1}_{js}$ both sides of \eqref{dotG} and sum over $j$ to obtain 
\bea
\dot \mathcal G \mathcal G^{-1} w_s + \mathcal G  \sum_j G^{-1}_{js}\dot v_j = \sum_{k} (\dot G G^{-1})_{ks} w_k + \dot w_s,
\label{motionpoles}
\eea
or equivalently 
\be
\label{motionpoles2}
 \sum_{k} (\dot G G^{-1})_{ks} w_k = \dot \mathcal G \mathcal G^{-1} w_s-  \dot w_s+ \mathcal G  \sum_j G^{-1}_{js}\dot v_j .
\ee
Tracing this last over $W$ on both sides gives the second statement. {\bf Q.E.D.}\\
For simplicity we shall consider now separately the two types of deformations, starting with those of $\Gamma$; we shall denote by $\pa$ any derivative (infinitesimal deformation or vector field on the parameter space $\un t$) of $\Gamma$.
\bp
Let $\pa $ denote any infinitesimal deformation of $\Gamma(z,\un t)$ and let $\S = \Gamma (z,\un t)D(z)$ where $D$ is as in \eqref{Ddef}. 
\label{proptraceable} Consider $\mathcal G= C_-[\bullet \S]$ as an operator $\H_-\to \H_-$. Then:\\
{\bf [1]} The operator below  is of finite rank and its range is $V=C_-[\H_+ \S ^{-1}]$:
\be
 \mathcal G^{-1} \pa \mathcal G +\mathcal  M _\pa  \label{traceable}\ ,\ \ \ \,\mathcal G:= C_-[\bullet  \S ];\ \  \mathcal  M_\pa  := C_-[\bullet \pa   \S   \S ^{-1}],
\ee
where $\pa \S \S^{-1} = \pa \Gamma \Gamma^{-1} \in \H_+$.\\
{\bf [2]} The  trace of \eqref{traceable} is 
\be
\tr \le(\mathcal G^{-1} \pa \mathcal G + \mathcal M_\pa\ri) = 
\oint_{\pa \D} \tr \le(R^{-1} R' \pa  \S   \S ^{-1} \ri) \frac {\d z}{2i\pi}.\label{taudiff}
\ee
\ep
\br
The expression $\tr \mathcal G^{-1} \pa \mathcal G$ would be the variation of the determinant of $\mathcal G$ if this operator had a determinant; in general, however,   $\mathcal G^{-1} \pa \mathcal G$ is   a T\"oeplitz operator on an infinite dimensional space and in particular not a trace-class perturbation of the identity, and thus there is no notion of determinant.
\er
{\bf Proof of Prop. \ref{proptraceable}}
{\bf [1]}
Let $h_-\in \H_-$. Using $\Id = C_+-C_-$ we have
\bea
&\& C_-[C_-[h_-\pa  \S ]  R_+^{-1} ]R + C_-[ h_-\pa  \S   \S ^{-1}]
 =-C_-[h_-\pa  \S ]  R_+^{-1}  R  + C_+[C_-[h_-\pa  \S ]  R_+^{-1} ]R   - h_-\pa  \S   \S ^{-1}  + C_+[h_-\pa  \S   \S ^{-1}]=\cr
\&\& =-C_+[h_-\pa  \S ]  \S ^{-1} + C_+[C_-[h_-\pa  \S ]  R_+^{-1} ]R  + C_+[h_-\pa  \S   \S ^{-1}].\nonumber
\eea
Since the left side is in $\H_-$, we take the projection $C_-$ (which only affects it by a minus sign) and thus 
\bea
C_-[C_-[h_-\pa  \S ]  R_+^{-1} ]R  + C_-[ h_-\pa  \S   \S ^{-1}] = C_-[C_+[h_-\pa  \S ]  \S ^{-1}] -   C_-[C_+ [C_-[h_-\pa  \S ]  R_+^{-1} ]R ]. \label{313}
\eea
The first term on the right side of \eqref{313} clearly belongs to $V = C_-[\H_+ \S^{-1}]$ already; as for the second term,  we know that $R  = \1 + L$ and that (the rows of) $L \in C_-[\H_+  \S ^{-1}]$ and then  we have 
$$
 C_-[\H_+R ] = C_-[\H_+C_-[\H_+ \S ^{-1}]] =C_-\le[- \H_+  \S ^{-1}  + \H_+ C_+ [\H_+ \S ^{-1}]\ri] = C_-[\H_+  \S ^{-1}].
$$
Therefore the trace is a finite-dimensional trace on $V = C_-[\H_+ \S ^{-1}]$.\\
{\bf [2]} The operator in \eqref{traceable} is  (using (\eqref{inverseG}) 
\bea
&\& C_-\le[C_-[h_-\pa  \S ] R_+^{-1}\ri]R   + C_-\le [h_-\pa  \S   \S ^{-1}\ri] =\nonumber\\
&\& =
- C_-\le[h_-\pa  \S  R_+^{-1}\ri]R   + \cancel{C_-\le[C_+ [h_-\pa  \S ] R_+^{-1}\ri]}R  + C_-\le[h_-\pa  \S   \S ^{-1}\ri]=\nonumber\\
&\& =- C_-\le[h_-\pa  \S   \S ^{-1} R ^{-1}\ri]R   + C_-\le[h_-\pa  \S   \S ^{-1}\ri].  \nonumber
\eea
Writing out the Cauchy projectors explicitly we find:
\bea
- C_-\le[h_-\pa  \S   \S ^{-1} R ^{-1}\ri]R   + C_-\le[h_-\pa  \S   \S ^{-1}\ri] = \oint_{\pa \D} \frac {\d \xi}{2i\pi} \oint_{\pa \D} \frac {\d \z}{2i\pi} \frac {h_-(\z) \pa  \S (\z)  \S ^{-1}(\z)\le(\1 -  R ^{-1}(\z)  R (\xi)\ri)}{(\z-\xi_-)(\xi-z_-) }.\nonumber
\eea
It is easily seen that the integration is independent of the order because the integrand is analytic at $\xi=\z$. 
 Moreover:
\be
\lim_{\z\to \xi} \frac {\1 -  R ^{-1}(\z)  R (\xi)}{\z-\xi } = 
\lim_{\z\to \xi} {  R ^{-1}(\z)  R '(\xi)} =   R ^{-1}(\z)  R '(\z).
\ee
It then follows that the trace is given by \eqref{taudiff}.
%%%%%%%%%%%%%%%%%%%%%%%%%%%%%%%%%%
 {\bf Q.E.D.}

\bc
\label{dlogdetG1}
Under the same assumptions and notation of Proposition  \ref{proptraceable} we have 
$$
\pa \ln \det G = \oint_{\pa \D} \frac {\d z}{2i\pi} \tr \le(R ^{-1} R ' \pa \S \S^{-1} \ri ) = \oint_{\pa \D} \frac {\d z}{2i\pi} \tr \le(R ^{-1} R ' \pa \Gamma\Gamma^{-1} \ri ).
$$
\ec
{\bf Proof.}
We have seen in Prop. \ref{proptraceable} that  
$$
\oint {\rm Tr} \le(R ^{-1} R ' \pa  \S   \S ^{-1}\ri) \frac { \d z}{2i\pi}  =\tr_{\H_-} \le(  \mathcal G^{-1} \pa \mathcal G + C_-[\bullet \pa  \S   \S ^{-1}] \ri) = 
\tr_{\H_-} \le(  \mathcal G^{-1} \pa \mathcal G +\mathcal  M _\pa \ri).
$$
We have also  seen in the same proposition that the trace is saturated on $V=C_-[\H_+\S ^{-1}]$, namely
\be
 \tr_{\H_-} \le(  \mathcal G^{-1} \pa \mathcal G +\mathcal  M _\pa\ri)  =  
 \tr_{V} \le(  \mathcal G^{-1} \pa \mathcal G+\mathcal  M _\pa\ri).
\ee
The basis \eqref{Wbasis} in $W$ is independent of the deformations of $\Gamma$; on the other hand the basis of $V$ \eqref{Vbasis} 
is obtained by choosing some vector $h_+(z)\in \H_+$ independent of the deformation and projecting it like $v = C_-[h_+ \S^{-1}]$ (with different choices of $h_+(z)$ as displayed in \eqref{Vbasis}, but always independent of deformations of $\Gamma$). Then 
\be
\pa v = C_-\le[h_+ \pa \S^{-1}\ri]  = -C_-\le[ h_+ \S^{-1} \pa \S \S^{-1}\ri].
\label{132}
\ee
Since $\pa \S\S^{-1}= \pa \Gamma \Gamma^{-1} \in \H_+$ then we can continue 
$$
\eqref{132} =  -C_-\le[\le(C_+[ h_+ \S^{-1}] - C_-[h_+\S^{-1}]\ri)  \pa \S \S^{-1}\ri] =  C_-\le[ C_-[h_+\S^{-1}]  \pa \S \S^{-1}\ri]
 = \mathcal M_\pa [v]\ .
$$
Thus
\bea
\oint {\rm Tr} \le(R ^{-1} R ' \pa  \S   \S ^{-1}\ri) \frac { \d z}{2i\pi}  \mathop{=}^{\tiny \tt Prop. \ref{proptraceable}_{\bf[2]}} \tr_{\H_-} \le(  \mathcal G^{-1} \pa \mathcal G +\mathcal  M _\pa\ri)  \mathop{=}^{\tiny \tt Prop. \ref{proptraceable}_{\bf[1]}} 
 \tr_{V} \le(  \mathcal G^{-1} \pa \mathcal G + \mathcal  M_\pa\ri)=\nonumber\\
  =\tr_V\le(\mathcal V_\pa\ri)\mathop{=}^{\tt Prop.\ref{lemmamoving}} \tr_{\C^m} \le(G^{-1}\pa G\ri) = \pa \ln \det G.\nonumber
\eea
{\bf Q.E.D.}\par \vskip 5pt

\br
The  major considerations we have made apply equally well (without almost any change) if we replace the locally defined diagonal rational function $D(z)$ by any locally defined (i.e. not necessarily diagonal) rational function $Q(z)$. In particular:
\begin{itemize}
\item The finite dimensionality of $V = C_-[\H_+ Q^{-1} \Gamma^{-1}], W = C_-[\H_+ \Gamma Q]$;
\item  as long as the total index of $\det Q(z)$ around $\D$ is zero, also Prop. \ref{propR} (with the obvious modification of \eqref{detR}), Prop. \ref{invertG}, Lemma \ref{lemmamoving}, Prop. \ref{proptraceable}, Cor. \ref{dlogdetG1} 
\item The choice of bases \eqref{Vbasis}, \eqref{Wbasis} would need modification (depending on the cases) but Lemma \ref{lemmamoving} is really of general nature and it is non-committal as to the specific choice of bases.\end{itemize}
The reason why this is interesting is because the process of ``soliton insertion'' in the literature  is accomplished by means of choosing a  $Q(z)$ of the form $\1 +N(z)$ with $N(z)$ a triangular rational matrix.  Only the part that follows after this remark (i.e. the dependence on the position of the poles) would require some more significant modification.
\er
We now turn our attention to the dependence of $G(\un t,\un c)$ on the positions of the points in $\mathcal A,\mathcal B$.
Suppose $\pa$ denotes the deformation of the position of a point $b\in \mathcal B$; we note that we shall consider {\em disjoint} union of $\mathcal A, \mathcal B$.  This means that if $z=a=b$ is at the same time a pole and a zero of $D(z)$ (clearly, in different entries along the diagonal!), we consider the deformation of the position of the pole  {\em independently} from that of the position of the zero.
\bt
\label{thmSchles}
Under any deformation $\pa$ of $\Gamma$, and of the position of poles/zeroes of $D$ we have 
\bea
\pa \ln \det G = \oint_{\pa \D} \tr \le(R ^{-1}R ' \pa   \S   \S ^{-1} \ri) \frac {\d z}{2i\pi} + \oint_{\pa \D} \tr \le(\Gamma^{-1}\Gamma' \pa  D D^{-1} \ri) \frac {\d z}{2i\pi}  \\
\S(z;\un t, \un c) = \Gamma(z;\un t) D(z;\un c)\nonumber 
\eea
where $G = G_{\le\{
\mathcal A \,\mathcal B\atop L\, K
\ri\}}$ is the characteristic matrix of Prop. \ref{characteristicmatrix} (and thus of Def. \ref{defGmatrix}).
\et
This has the immediate corollary 
\bc
\label{cor22}
The anomaly of the Malgrange differential under Schlesinger transformations is 
\be
\Omega\le(\pa; [M],
{\le\{{\mathcal A\, \mathcal B}\atop{ L\,K}\ri\}}\ri)  - \Omega\le(\pa; [M]\ri) =  \pa \ln \det G_{
{\le\{{\mathcal A\, \mathcal B}\atop{ L\,K}\ri\}}}.\ee
\ec
The proof of the corollary is simply the comparison of Thm. \ref{thmSchles} and Prop. \ref{propSchles}.

{\bf Proof of Thm. \ref{thmSchles}.}
The case of deformations of $\Gamma$ with fixed poles (i.e. with $\pa D\equiv 0$)  was already addressed in Cor.  \ref{dlogdetG1}  and thus we only need to address a deformation of the positions of the poles/zeroes of $D(z)$ fixing  $\Gamma$ ($ \pa\Gamma \equiv 0$).
We separate the proof in two cases, according to which type of pole we are moving (i.e. pole $b\in \mathcal B$ of $D$ or pole $a\in \mathcal A$ of $D^{-1}$).
\paragraph{Motion of poles $b\in\mathcal B$ of $D$:}
The basis \eqref{Vbasis} of $V$ is independent of $b\in \mathcal B$.
The basis \eqref{Wbasis}  is obtained by projection of elements of $\H_+$ of the form   $h_+^{(b, \mu, k)} = {\bf e}_\mu^tz_b^{K_b-\ell E_{\mu\mu}} \Gamma^{-1}(z)$ so $C_-[h_+^{(b, \mu, \ell)}  \S  ] ={\bf e}_\mu^tC_-[z_b^{-k}]$ 
and $0< k \leq (K_{b})_{\mu\mu}=k_{b,\mu}$.
  Clearly, the number of such vectors is $\tr K_b$. 
 Now, for any such vector (we drop the superscripts) $w = C_-[h_+ \S ]$, using \eqref{Wmove} we have:
\bea
\label{1231}
\mathcal W_\pa [w] &\&= \pa \mathcal G \mathcal G^{-1} [w] - \pa w   
= -C_-[C_-[h_+ R ^{-1}] R  \pa  \S ] -  C_-[h_+\pa   \S ]  - C_-[\pa h_+  \S ] = \\
&\& =  C_-[C_+ [h_+ R ^{-1} ] R_+ D \pa D]- C_-[\pa h_+  \S ]
\label{513}
\eea
We now need to compute the trace $\tr_{_W}\mathcal W_\pa$.
Looking at \eqref{513}, we observe that the last term, $C_{-} [\pa h_+ \Gamma D]$ is  a strictly triangular matrix and thus with zero trace; indeed  $\pa_b h_+^{(b,\mu,k)}   = k h_+^{(b,\mu,k-1)}$.
 Therefore the  only term that contributes to the trace is the first in \eqref{513}.
 Using \eqref{componentW} for the components along the chosen basis \eqref{Wbasis},  the trace is computed below
\bea
\tr(\pa  G G^{-1}) = \tr\eqref{1231}&\&= 
\sum_{\mu=1}^N \sum_{k = 1}^{k_{b, \mu}} \res_{z_b=0} C_-\le[C_+ \le[h_+^{(b, \mu, k)}  \S  R_+^{-1}\ri ] R_+ \pa_b D D^{-1}
\ri]\cdot {\bf e}_\mu z_b^{k} \frac {\d z_b}{z_b}  =\\
 \label{516}
&\&= \sum_{\mu=1}^N \sum_{k = 1}^{k_{b, \mu}} \res_{z_b=0}
\ C_-\le [C_+ \le [{\bf e}_\mu^t z_b^{-k} R_+^{-1}\ri ] R_+ \frac {K_b {\bf e}_\mu}{z_b}\ri ]z_b^k  \frac {\d z_b}{z_b} 
\eea
In the sum \eqref{516} clearly only the term $k=1$ is nonzero because the term $C_-[...]$ has at most a simple pole,  hence 
\bea
\eqref{516} 
&\& =  \sum_{\mu=1}^N  \res_{z_b=0}
\ C_-\le [C_+ \le [{\bf e}_\mu^t z_b^{-1} R_+^{-1}\ri ] R_+ \frac {K_b {\bf e}_\mu}{z_b}\ri ]  {\d z_b}=\nonumber\\
&\& = \res_{z=b}\tr\le( R_+^{-1} R_+' \frac {K_b}{z+b}\ri) = \oint_{\pa \D} \tr \le(R_+^{-1} R_+' \pa_b D D^{-1}\ri) \frac {\d z}{2i\pi} =\nonumber\\
&\& =\oint_{\pa \D} \tr \le( \S  ^{-1} R ^{-1} \le(R ' \S  + R   \S '\ri) \pa_b D D^{-1}\ri) \frac {\d z}{2i\pi}  = \nonumber\\
&\&
\oint_{\pa \D} \tr \le(R ^{-1}R ' \Gamma \pa_b D D^{-1} \Gamma^{-1}\ri) \frac {\d z}{2i\pi} + \oint_{\pa \D} \tr \le(\Gamma^{-1} \Gamma'  \pa_b D D^{-1}\ri) \frac {\d z}{2i\pi} +  \oint_{\pa \D} \tr \le(D^{-1}D'  \pa_b D D^{-1}\ri) \frac {\d z}{2i\pi}.  \nonumber
\eea
Note that the last term is zero because $D^{-1} D' \pa_b D D^{-1} = \frac {K_b}{{z_b}^2}$ for $z\in \D_a$ and $0$ in the other disks, and the residue vanishes. 
Thus we have 
\be
\tr \le(\pa_b GG^{-1}\ri) = \oint_{\pa \D} \tr \le(R ^{-1}R ' \Gamma \pa_b D D^{-1} \Gamma^{-1}\ri) \frac {\d z}{2i\pi} + \oint_{\pa \D} \tr \le(\Gamma^{-1} \Gamma'  \pa_b D D^{-1}\ri) \frac {\d z}{2i\pi}.
\label{DoneW} 
\ee
\paragraph{Motion of poles $a\in \mathcal A$ of $D^{-1}$}; this time the basis of $W$ is independent of the deformation, and thus we shall use Prop. \ref{lemmamoving} in the form of \eqref{Vmove}; we then need to find the action of $\mathcal V_\pa$.
The basis \eqref{Vbasis}  is obtained by projection of elements of $\H_+$ of the form   $h_+^{(a, \mu, \ell)} = {\bf e}_\mu^tz_a^{L_a-\ell E_{\mu\mu}}$ so $C_-[h_+^{(a, \mu, \ell)}  \S ^{-1}  ] ={\bf e}_\mu^tC_-[z_a^{-\ell} \Gamma^{-1}(z) ]$ 
and $0< \ell \leq (L_{a})_{\mu\mu}=\ell_{a,\mu}$.
  Clearly, the number of such vectors is $\tr L_a$. 
 For such a $v = C_-[h_+ \S ^{-1}]$ (we drop the superscripts) we have 
\bea
\mathcal V_\pa&\&= \mathcal G^{-1} \pa \mathcal G[v] + \pa v  =\mathcal G^{-1} \pa \mathcal G[v] - C_-[h_+  \S ^{-1} \pa  \S   \S ^{-1} ] + C_-[\pa h_+ \S ^{-1}]=\label{521}\\
&\&= \mathcal G^{-1} \pa \mathcal G[v] +  \mathcal  M_\pa [v] - C_-[C_+[h_+  \S ^{-1}]\Gamma \pa D D^{-1} \Gamma^{-1}]+ C_-[\pa h_+ \S ^{-1}]\label{524}
\eea
Looking at \eqref{524}, we observe that the last term, $C_{-} [\pa h_+ D^{-1}\Gamma ^{-1}]$ is  a strictly triangular matrix and thus with zero trace; indeed  $\pa_a h_+^{(a,\mu,\ell)}   = \ell  h_+^{(a,\mu,\ell-1)}$.

 Therefore the  only the first three terms in \eqref{524} contribute to the trace;  by Prop. \ref{proptraceable} of these, the first two are precisely \eqref{taudiff} and thus we need only to evaluate the trace of the third term. 
  Note that the component of any vector $v\in V$ along the element $v_{a,\mu,\ell}$ is computed by \eqref{componentV}
  and  thus the trace is computed 
\bea
\tr(\pa  G G^{-1}) \mathop{=}^{Lemma \ref{lemmamoving}} \tr_V\eqref{521}&\&=\nonumber\\
= \tr_{V} \big(\mathcal G^{-1} \pa_a \mathcal G + \mathcal  M_{\pa_a}\big) &\& 
-\sum_{\mu=1}^N \sum_{\ell = 1}^{\ell_{a, \mu}} \res_{z_a=0}
 C_-\le[C_+[h_+^{(a,\mu,\ell)}  \S ^{-1}]\Gamma \pa_a D D^{-1} \Gamma^{-1}\ri]  \Gamma(z) {\bf e}_\mu {z_a}^\ell \frac { \d z_a}{z_a} = \nonumber\\
\mathop{=}^{Prop. \ref{proptraceable}} 
\oint_{\pa \D} \tr \le(R ^{-1}R '  \pa_a \S\S^{-1}\ri) \frac {\d z}{2i\pi} 
&\& -\res_{z_a=0} \sum_{\mu=1}^N \sum_{\ell = 1}^{\ell_{a, \mu}} C_-\le[C_+\le[\frac {{\bf e}_\mu^t}{z_a^\ell}  \Gamma^{-1}\ri] \Gamma \frac {-L_a\Gamma^{-1}}{z_a} \ri] \Gamma{\bf e}_\mu{z_a}^{\ell } \frac {\d z_a}{z_a} \label{527}
\eea
Likewise before, in the sum \eqref{527} only the term $\ell=1$ is nonzero and since then the poles involved in the residue are all simple that term is computed as follows: 
\bea
 &\& -\res_{z_a=0} \sum_{\mu=1}^N  C_-\le[C_+\le[\frac {{\bf e}_\mu^t}{z_a}  \Gamma^{-1}\ri] \Gamma \frac {-L_a\Gamma^{-1}}{z_a} \ri] \Gamma{\bf e}_\mu {\d z_a} 
 =-\res_{z_a=0} \sum_{\mu=1}^N  C_+\le[\frac {{\bf e}_\mu^t}{z_a}  \Gamma^{-1}\ri] \Gamma \frac {-L_a\cancel{\Gamma^{-1}}}{z_a}\cancel{ \Gamma}{\bf e}_\mu {\d z_a} =\nonumber\\
 &\& 
 =-\res_{z_a=0} \sum_{\mu=1}^N  {\bf e}_\mu^t (\Gamma^{-1})'\bigg|_{z=a} \Gamma \frac {-L_a}{z_a}{\bf e}_\mu {\d z_a} 
  =  \res_{z=a} \tr \le( \Gamma^{-1} \Gamma'  \frac{-L_a}{z_a}\ri) = \oint_{\pa \D} \tr \le(\Gamma^{-1} \Gamma' \pa_a D D^{-1}\ri) \frac {\d z}{2i\pi},
\nonumber\eea
and hence 
\bea
\tr \le(G^{-1} \pa_a G\ri) = \oint_{\pa \D} \tr \le(R ^{-1}R '  \pa_a \S\S^{-1}\ri) \frac {\d z}{2i\pi} + \oint_{\pa \D} \tr \le(\Gamma^{-1} \Gamma'  \pa_a D D^{-1}\ri) \frac {\d z}{2i\pi}.
\label{DoneV}
\eea
This completes the proof. {\bf Q.E.D.} \par \vskip 5pt
The Theorem \ref{thmSchles} and Cor. \ref{cor22} are the main ingredient of the proof of Thm. \ref{maintheorem}: however in Thm. \ref{maintheorem} the diagonal matrix is {\em globally defined} and not piecewise.  We need to see which additional term arises in Thm. \ref{thmSchles} if $D(z)$ is replaced by a globally defined diagonal rational function with the same zeroes/poles.

We shall now denote by $D_{loc}(z)$ the diagonal matrix that was denoted by $D$ thus far. 
\bt
\label{thmSchlesglobal}
Let  $R(z) = R_{\le\{{\mathcal A\,\mathcal B}\atop{L\,K}\ri\}}(z)$ be Schlesinger transformation of Def. \ref{defSchles}. Let $\wt \S = \Gamma(z) D(z)$ and $D(z)$ the diagonal rational function $D(z) =  \prod_{c\in \mathcal B\cup \mathcal A} z_c^{L_c-K_c}$. 
If $\pa$ denotes any deformation of $\Gamma$ or of the zeroes/poles of $D$ then  
\be
\pa \ln \det G =  \oint_{\pa \D} \tr \le(R ^{-1}R ' \pa   \wt \S  \wt  \S ^{-1} \ri) \frac {\d z}{2i\pi} + \oint_{\pa \D} \tr \le(\Gamma^{-1}\Gamma' \pa  D D^{-1} + D^{-1}D' D^{-1} \pa D \ri) \frac {\d z}{2i\pi}.
\ee
\et
{\bf Proof.} 
To be clear, $D(z)$ and $D_{loc}$ are related by ($\chi_{_S}$ denotes the indicator function of a set $S$) 
$$
D(z) = \prod_{c\in \mathcal A\cup \mathcal B} z_c^{L_c-K_c}\ ,\ \ \ D_{loc} (z) = \prod_{c\in \mathcal A\cup \mathcal B} z_c^{(L_c-K_c)\chi_{_{\D_c}}}\ ,\ \ T(z) = D(z)/D_{loc}(z);
$$
so that, within a given disk $\D_{c_0}$ the matrix $T(z)$ is analytic and given by $
T(z) = \prod_{c; c\neq c_0}  z_c^{L_c-K_c},$ $z\in \D_{c_0}.$
Let $\pa$ be a derivative of the position of any zero/pole of $D(z)$ (so $\pa \Gamma \equiv 0)$, then\footnote{
When differentiating $D_{loc}(z)$ w.r.t. the position of a pole, we understand it for fixed $\D$; for example if $b$ is a point of $\mathcal B$ then $D_{loc}^{-1} \pa_b D_{loc} = \frac {K_b}{z-b} \chi_{\D_b}$.
}
\bea
&\& \oint_{\pa \D} \tr \le(\wt R ^{-1} \wt R ' \pa \wt \S \wt \S^{-1}\ri) \frac {\d z}{2i\pi}
 =
\sum_{c\in \mathcal A \cup \mathcal B} \res_{z=c} \tr \le(R ^{-1} R ' \Gamma \pa  D  D^{-1} \Gamma^{-1} \ri)  {\d z}= 
\nonumber \\
&\& =
\sum_{c\in \mathcal A \cup \mathcal B} \res_ {z=c} \tr \le(R ^{-1} R ' \Gamma \pa  D_{loc}  D_{loc}^{-1} \Gamma^{-1} \ri)  {\d z} + 
\sum_{c\in \mathcal A \cup \mathcal B} \res_ {z=c} \tr \le(R ^{-1} R ' \Gamma D_{loc} \pa T T^{-1} D_{loc}^{-1} \Gamma^{-1} \ri)  {\d z}= 
\nonumber \\
&\& =
\oint_{\pa \D} \tr \le(R ^{-1} R ' \pa  \S \S ^{-1}\ri) \frac {\d z}{2i\pi} + 
\sum_{c\in \mathcal A \cup \mathcal B} \res_ {z=c} \tr \le(R_+^{-1} R ' \S  \pa T T^{-1} \ri)  {\d z}\label{step1}
\eea
where $\S = \Gamma D_{loc}$, namely,  it has exactly the same meaning as in Thm. \ref{thmSchles}.
Use now that $R_+ :=R \S $ is analytic;  then  plug  $R  = R_+ \S^{-1} = R_+ D_{loc}^{-1} \Gamma^{-1}$ into \eqref{step1} and simplify (using the cyclicity of the trace) to obtain
\bea
\oint_{\pa \D}&\&  \tr \le(R ^{-1} R ' \pa  \S  \S^{-1}\ri) \frac {\d z}{2i\pi} + \cr
&\& +
\sum_{c\in \mathcal A \cup \mathcal B} \res_ {z=c} \bigg[
\overbrace{\tr \le(R_+^{-1} R_+' \pa T T^{-1} \ri) }^{\hbox{ analytic $\Rightarrow$ no residue}}
-
\tr \le(D_{loc}^{-1} D_{loc}' \pa T T^{-1} \ri) 
-
\overbrace{\tr \le(\Gamma^{-1}\Gamma'   \pa T T^{-1} \ri)}^{\hbox{ analytic $\Rightarrow$ no residue} }
 \bigg] {\d z}= \cr
&\& =\oint_{\pa \D} \tr \le(R ^{-1} R ' \pa  \S  \S ^{-1}\ri) \frac {\d z}{2i\pi} 
-
\sum_{c\in \mathcal A \cup \mathcal B} \res_ {z=c} 
\tr \le(D_{loc}^{-1} D_{loc}' \pa T T^{-1} \ri)  {\d z}.\nonumber
\eea
Therefore the two relative $\tau$ differentials differs by the explicit term 
\be
\oint_{\pa \D} \tr \le(R ^{-1} R ' \pa  \wt \S \wt  \S^{-1}\ri) \frac {\d z}{2i\pi} - \oint_{\pa \D} \tr \le(R ^{-1} R ' \pa  \S  \S^{-1}\ri) \frac {\d z}{2i\pi}
=
- \sum_{c\in \mathcal A \cup \mathcal B} \res_ {z=c} 
\tr \le(D_{loc}^{-1} D_{loc}' \pa T T^{-1} \ri).\label{1323}
\ee
It is easy to see that the  right side of \eqref{1323}   is  expressed as the sum of the residues of $\tr (D'D^{-1} \pa D D^{-1})$. 
{\bf Q.E.D.}
We thus can finally prove Theorem \ref{maintheorem}
\subsubsection*{Proof of Thm. \ref{maintheorem}}
\label{proofmain}
{\bf [1]} This point is obvious from the very definition of $R$ and simple computations.\\
{\bf [2]} 
We compute the integrand in $\omega(\pa; [\wt M])$. To this end we use the following identities
\bea
\wt \Gamma^{-1} \wt \Gamma' &\& =  D^{-1}\Gamma^{-1}  R^{-1} R' \Gamma D +  D^{-1}  \Gamma^{-1} \Gamma' D + D^{-1} D' ,
\nonumber\\
\pa (D^{-1} M D) D^{-1}M^{-1} D &\& =  D^{-1} \pa M M^{-1} D - \pa D D^{-1} + D^{-1}MD^{-1} \pa D M^{-1} D,\nonumber\\
\label{ww}
\Gamma_- \pa M M^{-1} \Gamma_-^{-1} &\& =\pa  \Gamma_+ \Gamma_+^{-1} - \pa \Gamma_- \Gamma_-^{-1}\ .
\eea
Therefore, plugging in and simplifying (using the cyclicity of the trace several times and \eqref{ww}), we find (below $\Delta$ denotes the jump operator $\Delta (f) = f_+ - f_-$):
\bea
&\tr& \bigg[
\wt \Gamma_-^{-1}  \wt \Gamma'_-  \pa (D^{-1} M D) D^{-1}M^{-1} D
\bigg]=\nonumber\\
&=& \tr \bigg[
\Gamma_-^{-1} \Gamma_-' \pa M M^{-1} + 
R^{-1} R'  {\Delta \le(\pa (\Gamma D) D^{-1}\Gamma^{-1}\ri)} + 
{ \Delta \le( \Gamma^{-1} \Gamma' \pa DD^{-1} \ri)}+\nonumber\\
& 
-&{  M^{-1} M' \pa D D^{-1}} + D^{-1} D' \le(  
\pa M M^{-1}  - \pa D D^{-1} + MD^{-1} \pa D M^{-1} \ri)\bigg]. \nonumber
\eea
Now, if we have $\int_\Sigma \Delta F \frac{\d z}{2i\pi}$ and $F$ has some poles outside of $\Sigma$ then this reduces, by the Cauchy theorem, to the sum of the residues of $F$. This concludes the proof of the second part. \\
{\bf [3]} From Theorem \ref{thmSchlesglobal} 
$$
\sum_{c} \res_{z=c} \tr \le(
R^{-1} R' \pa (\Gamma D) D^{-1} \Gamma^{-1} 
\ri) +
\res_{z=c} \tr \le(
\Gamma^{-1}\Gamma' \pa D D^{-1}\ri) = \pa \ln \det G  - \sum_{c} \res_{z=c}\tr \le( D^{-1}D' D^{-1} \pa D \ri) \d z   
$$
The Vandermonde-like expression is simply the explicit evaluation of the last term. Note that none of the terms has residue at $\infty$, which explains the exclusion in the final formula.
 This concludes the proof. {\bf Q.E.D.}

\bibliographystyle{plain}
\bibliography{/Users/mattiacafasso/Documents/BibDeskLibrary.bib}
 
 \end{document}